\documentclass{ws-jai}

\usepackage[flushleft]{threeparttable}
\usepackage{multirow}
\usepackage{comment}

\def\epic{{\sc Epic}}

\def\sk1p{{\sc Skip}}

\begin{document}

\catchline{}{}{}{}{} % Publisher's Area please ignore

\markboth{Bradley R. Johnson}{A CubeSat for Calibrating Ground-Based
  and Sub-Orbital Millimeter-Wave Polarimeters (CalSat)}

\title{A CUBESAT FOR CALIBRATING GROUND-BASED AND SUB-ORBITAL
  MILLIMETER-WAVE POLARIMETERS (CALSAT)}

\author{Bradley~R.~Johnson$^1$, Clement~J.~Vourch$^2$, Timothy~D.~Drysdale$^2$,
  Andrew Kalman$^3$, \\ Steve Fujikawa$^{4}$, Brian Keating$^{5}$ and Jon Kaufman$^{5}$}

\address{
$^1$Department of Physics, Columbia University, New York, NY 10027, USA \\
$^2$School of Engineering, University of Glasgow, Glasgow, Scotland G12 8QQ, UK \\
$^3$Pumpkin, Inc., San Francisco, CA 94112, USA \\
$^4$Maryland Aerospace Inc., Crofton, MD 21114, USA \\
$^5$Department of Physics, University of California, San Diego, CA 92093-0424, USA
}

\maketitle

\footnotetext[1]{bjohnson@phys.columbia.edu}

%%%%%%%%%%%%%%%%%%%%%%%%%%%%%%%%%%%%%%%%%%%%%%%%%%%%%%%%%%%%%%%%%%%%%%%%%%%%

\begin{history}
\received{(to be inserted by publisher)};
\revised{(to be inserted by publisher)};
\accepted{(to be inserted by publisher)};
\end{history}

%%%%%%%%%%%%%%%%%%%%%%%%%%%%%%%%%%%%%%%%%%%%%%%%%%%%%%%%%%%%%%%%%%%%%%%%%%%%

\begin{abstract}

We describe a low-cost, open-access, CubeSat-based calibration
instrument that is designed to support ground-based and sub-orbital
experiments searching for various polarization signals in the cosmic
microwave background (CMB).
All modern CMB polarization experiments require a robust calibration
program that will allow the effects of instrument-induced signals to
be mitigated during data analysis.
A bright, compact, and linearly polarized astrophysical source with
polarization properties known to adequate precision does not exist.
Therefore, we designed a space-based millimeter-wave calibration
instrument, called CalSat, to serve as an open-access calibrator, and
this paper describes the results of our design study.
The calibration source on board CalSat is composed of five ``tones''
with one each at 47.1, 80.0, 140, 249 and 309~GHz.
The five tones we chose are well matched to (i) the observation
windows in the atmospheric transmittance spectra, (ii) the spectral
bands commonly used in polarimeters by the CMB community, and (iii)
The Amateur Satellite Service bands in the Table of Frequency
Allocations used by the Federal Communications Commission.
CalSat will be placed in a polar orbit allowing visibility from
observatories in the Northern Hemisphere, such as Mauna~Kea in Hawaii
and Summit Station in Greenland, and the Southern Hemisphere, such as
the Atacama Desert in Chile and the South Pole.
CalSat also will be observable by balloon-borne instruments launched
from a range of locations around the world.
This global visibility makes CalSat the only source that can be
observed by all terrestrial and sub-orbital observatories, thereby
providing a universal standard that permits comparison between
experiments using appreciably different measurement approaches.

\end{abstract}


%%%%%%%%%%%%%%%%%%%%%%%%%%%%%%%%%%%%%%%%%%%%%%%%%%%%%%%%%%%%%%%%%%%%%%%%%%%%

\section{Introduction}
\label{sec:introduction} 

%%%%%%%%%%%%%%%%%%%%%%%%%%%%%%%%%%%%%%%%%%%%%%%%%%%%%%%%%%%%%%%%%%%%%%%%%%%%

In this paper, we describe a low-cost, open-access, CubeSat-based
calibration instrument called CalSat that is designed to support
ground-based and sub-orbital experiments searching for various
polarization signals in the cosmic microwave background (CMB).
The CMB is a bath of photons that permeates all of space and carries
an image of the Universe as it was 380,000 years after the Big Bang.
This image spans the entire sky, but it is not visible to the human
eye because the frequency spectrum of the CMB peaks in the
millimeter-wave region of the electromagnetic spectrum.
Physical processes that operated in the universe when the CMB formed
left an imprint that can be detected today.
This imprint is observed as angular intensity and linear polarization
anisotropies.
These primordial CMB anisotropies have proven to be a treasure trove
of cosmological information.
For example, the precise characterization of the intensity (or
temperature) anisotropy of the CMB has helped reveal that spacetime is
flat, the universe is 13.8 billion years old, and the energy content
of the universe is dominated by cold dark matter and dark
energy~\citep[see for example,][]{bennett_2013,planck_2014}.
The density inhomogeneities that produced the detected temperature
anisotropies theoretically should generate, through Thomson scattering
during the epoch of recombination, a curl-free polarization signal
known as ``E-mode'' polarization.
This companion signal has also been observed at the theoretically
expected level providing further confidence in the $\Lambda$CDM
cosmological model~\citep[see for
  example,][]{naess_2014,crites_2015,bicep2i_2014,quiet_2011,quad_2009}.

%%%%%%%%%%%%%%%%%%%%%%%%%%%%%%%%%%%%%%%%%%%%%%%%%%%%%%%%%%%%%%%%%%%%%%%%%%%%

\begin{figure}[t]
\centering
\includegraphics[width=0.8\textwidth]{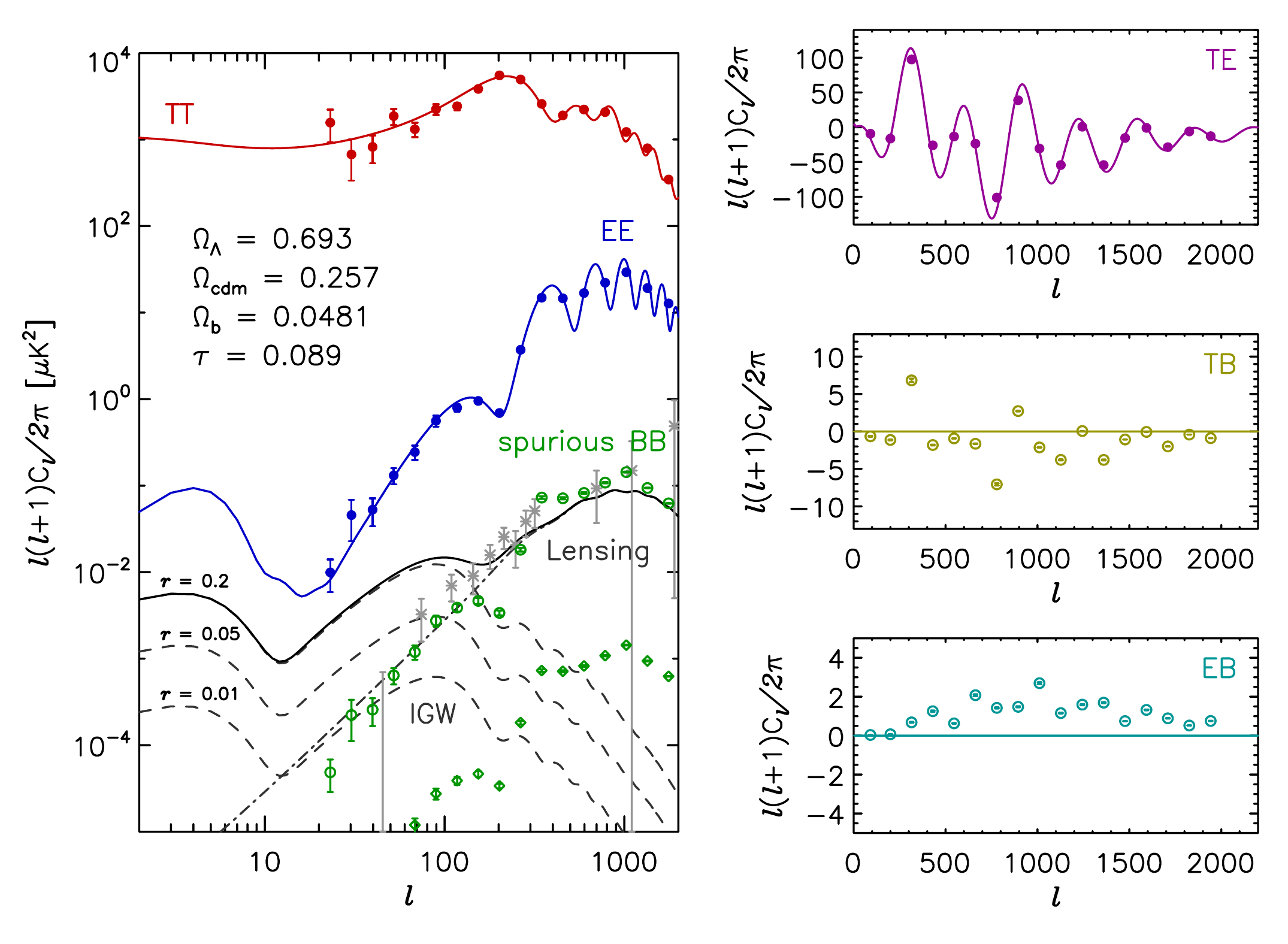}
\caption{Angular power spectra for various CMB signals.
The solid curves are models computed using the current best-fit
cosmological parameters.
Data points from an experiment simulation (circles and diamonds) and
measurements from the~\citet{polarbear_2014} and
the~\citet{bicep2_keck_planck_2015} (grey stars) are plotted along
with these curves for comparison.
On the left, the red curve corresponds to the temperature signal, the
blue curve corresponds to the E-mode signal, and the black curve is
the expected B-mode signal, which includes both an inflationary
gravitational wave (IGW) component (dashed) for $r = 0.2$ and a
non-primordial B-mode signal that is produced when large-scale
structure converts E-modes to B-modes via gravitational lensing
(dash-dot).
For comparison, IGW signal curves for $r = 0.01~\mbox{and}~0.05$ are also
plotted.
The TE, TB and EB power spectra are plotted on the right.
The details of the experiment simulation, which shows the effect of
instrument-induced polarization rotation, are discussed in
Section~\ref{sec:polarimeter_calibration}.
The green circles and diamonds show the spurious BB signal for 2.0~deg
and 0.2~deg of polarization rotation, respectively.}
\label{fig:ps}
\end{figure}

%%%%%%%%%%%%%%%%%%%%%%%%%%%%%%%%%%%%%%%%%%%%%%%%%%%%%%%%%%%%%%%%%%%%%%%%%%%%

The inflationary cosmological paradigm posits that a burst of
exponential spacetime expansion, called inflation, took place during
the first fraction of a second after the Big Bang.
Observational evidence to date supports this paradigm and has given it
a strong footing, though the precise physical mechanism that caused
inflation is unknown.
Inflation should have produced a stochastic background of
gravitational waves.
These gravitational waves would have produced polarization signals
separable from E-modes by their divergence-free ``B-mode''
signature~\cite{zaldarriaga_1997,kamionkowski_1997}.
The magnitude of this inflationary gravitational wave (IGW) B-mode
signal is proportional to the energy scale at which inflation
occurred.
See for example, the dashed curves in Figure~\ref{fig:ps} and note
that the IGW signal amplitude is commonly parameterized as $r$, the
so-called tensor-to-scalar ratio, which is related to the energy scale
of inflation\footnote{The tensor-to-scalar ratio $r$ is related to the
  energy scale of inflation as $\displaystyle V^{1/4} = 1.06 \times
  10^{16}
  \left(\frac{r}{0.01}\right)^{1/4}$~[GeV].}~\cite{baumann_2009,knox_2002}.
If the IGW signal ultimately is discovered, then the energy scale of
inflation would be experimentally ascertained.
This measurement would place a tight constraint on the theoretical
models that describe the inflation mechanism, and this constraint
would be a breakthrough for both astrophysics and particle physics
because there is no way to create inflation-like conditions in a
laboratory or particle accelerator.
The IGW signal is currently the primary focus of CMB research, and we
will discuss the status of current measurements at the end of
Section~\ref{sec:polarimeter_calibration}.

In addition to the IGW signal, other CMB polarization signals promise
to deliver valuable cosmological information.
Non-primordial B-modes are generated when E-modes are gravitationally
lensed by large-scale structures in the Universe (see the dash-dot
curve in Figure~\ref{fig:ps}).
This lensing B-mode signal is sensitive to physical parameters such as
the sum of the neutrino masses~\cite{abazajian_2015a}.
B-mode polarization can also be used to probe physics outside the
standard cosmological picture.
Temperature to B-mode correlations (TB) and E-mode and B-mode
correlations (EB) are expected to vanish in the standard model.
Therefore, these estimators are sensitive probes for physics outside
the standard model if the correlation signals are found to be
non-zero.
A variety of candidate non-standard-model physical mechanisms that
produce cosmological polarization rotation (CPR) already have been
identified.
Parity violation in the electromagnetic sector via a Chern-Simons
coupling can produce TB and EB correlations~\cite{kaufman_2014}.
A coupling of a pseudo-scalar field to electromagnetism would violate
the Einstein Equivalence Principle and would result in TB and EB
correlations~\cite{ni_1977,carroll_1991}.
TB and EB correlations can also be used to test chiral gravity
models~\cite{gluscevic_2010} and to search for primordial magnetic
fields~\cite{pogosian_2009,yadav_2012}.

%%%%%%%%%%%%%%%%%%%%%%%%%%%%%%%%%%%%%%%%%%%%%%%%%%%%%%%%%%%%%%%%%%%%%%%%%%%%

\begin{table}[t]
\small
\centering
\begin{threeparttable}
\begin{tabular}{|c|c|ccc|cc|cc|}
\hline
& frequency & $I$  & $P$  & $P/I$ & \multicolumn{2}{c|}{polarization angle uncertainty} & total power\tnote{$\dagger$} & polarized power\tnote{$\dagger$} \\
& [GHz]     & [Jy] & [Jy] & [\%]  & systematic [deg] & statistical [deg]                & [fW]        &  [fW]           \\
\hline
\multirow{7}{*}{Planck\tnote{$a$}} & 30 & 340 & 24 & 7.1 & 0.50 & 0.54 & 16 & 2.2 \\
 &  44 & 290 & 19 & 6.5 & 0.50 & 0.32 & 20 & 2.5 \\
 &  70 & 260 & 21 & 7.9 & 0.50 & 0.23 & 30 & 4.4 \\
 & 100 & 220 & 16 & 7.2 & 0.62 & 0.11 & 35 & 4.8 \\
 & 143 & 170 & 12 & 7.2 & 0.62 & 0.13 & 39 & 5.1 \\
 & 217 & 120 & 10 & 8.1 & 0.62 & 0.12 & 42 & 6.5 \\
 & 353 & 82  & 10 & 12  & 0.62 & 0.37 & 49 & 11  \\
\hline
\multirow{5}{*}{WMAP\tnote{$b$}} & 23 & 380 & 27 & 7.1 & 1.5 & 0.10 & 14 & 1.9 \\
 & 33  & 340 & 24 & 6.9 & 1.5 & 0.10 & 18 & 2.4 \\
 & 41  & 320 & 22 & 7.0 & 1.5 & 0.20 & 21 & 2.7 \\
 & 61  & 280 & 19 & 7.0 & 1.5 & 0.40 & 27 & 3.5 \\
 & 93  & 230 & 17 & 7.1 & 1.5 & 0.70 & 34 & 4.7 \\
\hline
IRAM\tnote{$c$} & 90 & 200 & 15 & 8.8 & 0.50 & 0.20 & 29 & 4.7 \\
\hline
\multirow{5}{*}{CalSat} & 47.1 & & & 100 & 0.05 & & 2000 & 2000 \\
 & 80.0 & & & 100 & 0.05 & & 1300 & 1300   \\
 & 140  & & & 100 & 0.05 & & 130  & 130  \\
 & 249  & & & 100 & 0.05 & & 48   & 48 \\
 & 309  & & & 100 & 0.05 & & 23   & 23 \\
\hline
\end{tabular}
\begin{tablenotes}
\item[$a$]{Measurements from~\citet{planck_2015} using the maximum likelihood filtering method}
\item[$b$]{Measurements from~\citet{weiland_2011}}
\item[$c$]{Measurement from~\citet{aumont_2010}}
\item[$\dagger$]{Computed for this comparison assuming the detector is
  sensitive to a single polarization and the telescope aperture area
  is 1~m$^2$ for all cases.  To convert the measurements from Planck,
  WMAP and IRAM from Jy to fW, we made the additional assumption that
  $\delta\nu/\nu_c = 0.3$}
\end{tablenotes}
\caption{
A comparison between CalSat and Tau~A measurements from Planck, WMAP
and the IRAM telescope.
The polarization angle uncertainty requirement from the Weiss Report
is 0.2~deg~\cite{weiss_2006}.
The available measurements of Tau~A do not meet this specification
because the smallest systematic error to date is 0.5~deg.
Also, the high-resolution~\citet{aumont_2010} measurement shows that
the polarization intensity morphology of Tau~A is complicated.
}
\label{table:tau_a}
\end{threeparttable}
\end{table}

%%%%%%%%%%%%%%%%%%%%%%%%%%%%%%%%%%%%%%%%%%%%%%%%%%%%%%%%%%%%%%%%%%%%%%%%%%%%

\section{Polarimeter Calibration}
\label{sec:polarimeter_calibration}

%%%%%%%%%%%%%%%%%%%%%%%%%%%%%%%%%%%%%%%%%%%%%%%%%%%%%%%%%%%%%%%%%%%%%%%%%%%%

The magnitude of the forecasted TB, EB and B-mode signals is faint
when compared with unavoidable instrument-induced systematic errors.
Therefore, all experiments trying to measure these signals need a
robust polarimeter calibration program that will allow the effect of
these instrument-induced errors to be mitigated during data analysis.
Several performance requirement studies have been published
specifically for IGW searches~\cite{weiss_2006,odea_2007,bock_2009}.
These studies indicate the level of systematic error that can be
tolerated for a given IGW signal amplitude.
For the CalSat design program we chose to use the performance
requirements derived by the Task Force on Cosmic Microwave Background
Research~\cite{weiss_2006} as a benchmark.
This report proposes a general performance requirement where each
systematic error must be suppressed to a factor of ten below an IGW
signal corresponding to a tensor-to-scalar ratio of $r = 0.01$, which
is the level where the gravitational lensing signal starts to dominate
the IWG signal at $\ell = 100$.

Many of the known systematic errors can be mitigated
straightforwardly.
However, a \textit{robust} mitigation strategy for instrument-induced
polarization rotation (IPR), which is one of the most critical
systematic errors, has not yet been identified.
This instrumental effect rotates the apparent orientation of the
linearly-polarized pseudo-vectors on the sky, and this rotation
converts the brighter E-modes into spurious B-modes.
Using the aforementioned performance requirement definition, any IPR
must be limited to 0.2~deg for an IGW search targeting $r > 0.01$;
fainter IGW signals would require a tighter constraint, and even more
precision could be needed if delensing techniques ultimately will be
used to probe below $r = 0.01$~\cite{smith_2012}.

To illustrate this effect we performed a simple experiment simulation
where a set of Stokes parameter maps ($I$, $Q$, $U$ with $V=0$)
containing CMB signals were simulated and then the orientation of the
polarization of each pixel was rotated by 2~deg using the Mueller
matrix for linear polarization rotation.
The input maps were 1000~deg$^2$, contained no noise or other beam
effects, and all B-mode signals were set to zero.
The angular power spectra of the corrupted maps were then estimated,
and the estimated power spectra are shown in Figure~\ref{fig:ps}.
The open-circle points in this figure (BB, TB and EB) are spurious
signals entirely produced by the polarization rotation operation.
A rotation of just 2~deg creates a false B-mode signal that is
comparable in magnitude to both the lensing signal and the
sought-after IGW signal below $\ell \simeq 100$ if the
tensor-to-scalar ratio is $r=0.01$.
The takeaway message is a very small amount of rotation can produce a
false B-mode signal that can obscure the actual sky signal.

To properly characterize the IPR properties of a CMB polarimeter, a
linearly polarized source with known polarization properties should be
observed with high precision.
This required calibration measurement provides the critical
relationship between the coordinate system of the polarimeter in the
instrument frame and the coordinate system on the sky that defines the
astrophysical $Q$ and $U$ Stokes parameters.
The ideal source should be point-like in the beam of the telescope and
bright enough to generate high signal-to-noise calibration data with a
reasonable amount of integration time.
But this source should not be too bright, or it will saturate the
detector system.
Finally, the orientation of the polarization of the observed source on
the sky must be known to 0.2~deg or better for $r > 0.01$.
It is important to note that the magnitude of this systematic error
typically varies across the focal plane of the telescope, so the IPR
effect must be determined for each detector in any given instrument.

Many current experiments ``self-calibrate'' their polarization angles
by assuming TB and EB correlations are zero~\cite{keating_2013}.
This self-calibration technique uses the TB and EB spectra on the
right in Figure~\ref{fig:ps} along with the theoretical idea that
these spectra should be zero in the absence of CPR to remove the green
spurious B-mode points on the left.
This approach makes it impossible to use B-mode, TB and EB
measurements to constrain the aforementioned isotropic departures from
the standard model (see Section~\ref{sec:introduction}) because any
signal produced by CPR is incorrectly interpreted as an error signal.
If this problem is avoided and self-calibration is not used, then the
current achieved precision of $\sim$0.5~deg severely limits the
ability of experiments to search for any departures from standard
cosmology via TB and EB measurements.
A more precise calibration procedure is needed to make progress with
CPR studies.

Some partially polarized astrophysical sources are available, and at
millimeter-wavelengths, the best of these appears to be Tau~A.
Tau~A is a supernova remnant located at Right ascension = 05~34~32
(hms) and Declination = +22~00~52 (dms), and it has an angular extent
of $7^{\prime}\times5^{\prime}$.
For calibration, its most relevant properties are (i) the measured
polarization angle uncertainty, (ii) the polarized fraction, and (iii)
the source brightness.
The WMAP satellite measured its polarization properties in spectral
bands between 20 and 100~GHz~\cite{weiland_2011}, the Planck satellite
measured its polarization properties in spectral bands between 30 and
353~GHz~\cite{planck_2015}, and \citet{aumont_2010} studied the
suitability of Tau~A as a calibration source for CMB studies at 90~GHz
using the 30~m IRAM telescope.
These measurements are summarized in Table~\ref{table:tau_a}.
ACTpol~\cite{naess_2014} and POLARBEAR~\cite{polarbear_2014} recently
observed Tau~A as part of their respective calibration programs and
showed that the polarization intensity morphology at 145~GHz is
complicated.

Though Tau~A is the best polarized source on the sky at millimeter
wavelengths, it does not meet the performance requirements of a
calibrator for an IGW-signal search.
First and foremost, with a Declination = +22~00~52 (dms), it is not
accessible to observatories at the South Pole or on Antarctic
balloons.
It can be observed from the Atacama Desert in Chile, though it will
always be at a zenith angle of 45~deg or more.
Second, it is not bright enough to give a high signal-to-noise ratio
measurement with a short integration time.
Third, the source is extended rather than point-like.
And finally, the millimeter-wave spectrum of Tau~A is not precisely
known, though it is certainly not that of a 2.7~K blackbody.
Polarimeters that have frequency-dependent performance, which is
common, would need to precisely measure the frequency spectrum of
Tau~A before using it as a polarization angle calibrator (see
Section~\ref{sec:detectability}).

Since an ideal \textit{celestial} calibration source does not exist,
some current experiments use ground-based sources for polarimeter
calibration~\cite{bicep2ii_2014,crites_2015}.
However, these measurements are challenging.
For example, the Fraunhofer distance ($2D^2/\lambda$) for common
telescopes is typically between 0.1 and 100~km, so the calibration
source must be placed more than 0.1 to 100~km away, respectively, from
the telescope for the desired far-field beam characterization
measurement.
This kind of measurement is difficult because the detector systems in
CMB polarimeters are designed to observe sky backgrounds, which are
typically $\sim$10~K or less.
To observe a ground-based calibration source more than 0.1 to 100~km
away, the zenith angle of the telescope must be set to nearly 90~deg.
Background loading from the large airmass and thermal emission from
the ground at this zenith angle are high enough to saturate typical
detector systems\footnote{Here we are assuming bolometric detectors
  are being used. Bolometers are commonly used in CMB observations.}.
Therefore two-stage bolometers or some kind of signal attenuator must
be used for this style of calibration measurement.
However, these approaches have their own varieties of systematic
error, so awkwardly these calibration measurements could require their
own calibration.
A near-field measurement of a calibration source observed at smaller
zenith angles can be performed instead to mitigate this high
background loading problem.
Yet, for the near-field measurement to be useful, a model-dependent
theoretical correction must be applied to the data to derive the
effective far-field calibration.
This correction, however, also introduces uncertainty into the
calibration, and the quality of the result depends entirely on the
accuracy of this difficult theoretical calculation.

%%%%%%%%%%%%%%%%%%%%%%%%%%
%% current measurements %%
%%%%%%%%%%%%%%%%%%%%%%%%%%

Current experiments are beginning to measure B-modes.
The BICEP2 Collaboration recently announced a detection of
degree-scale B-mode polarization in a 350~deg$^2$ patch of sky in the
southern hemisphere~\cite{bicep2i_2014}.
The POLARBEAR and SPTPol Collaborations detected B-mode polarization
consistent with the aforementioned gravitational lensing signal on
sub-degree scales~\cite{polarbear_2014,keisler_2015}.
And the Planck Collaboration published a measurement of B-mode
polarization from Galactic dust emission at 353~GHz~\cite{adam_2014}.
This combination of results has added energy to CMB studies because
these measurements indicate that faint B-mode signals are there on the
sky and that instruments are now sensitive enough to detect them.
Currently, BICEP2 and POLARBEAR use the aforementioned
self-calibration technique, so it impossible to use these measurements
to constrain isotropic departures from the standard model.
These results have re-emphasized the need for enhanced foreground
discrimination capabilities and robust systematic error control; the
latter effect being the major motivation behind CalSat.

%%%%%%%%%%%%%%%%%%%%%%%%%%%%%%%%%%%%%%%%%%%%%%%%%%%%%%%%%%%%%%%%%%%%%%%%%%%%

\begin{figure}[t]
\centering
\includegraphics[width=0.9\textwidth]{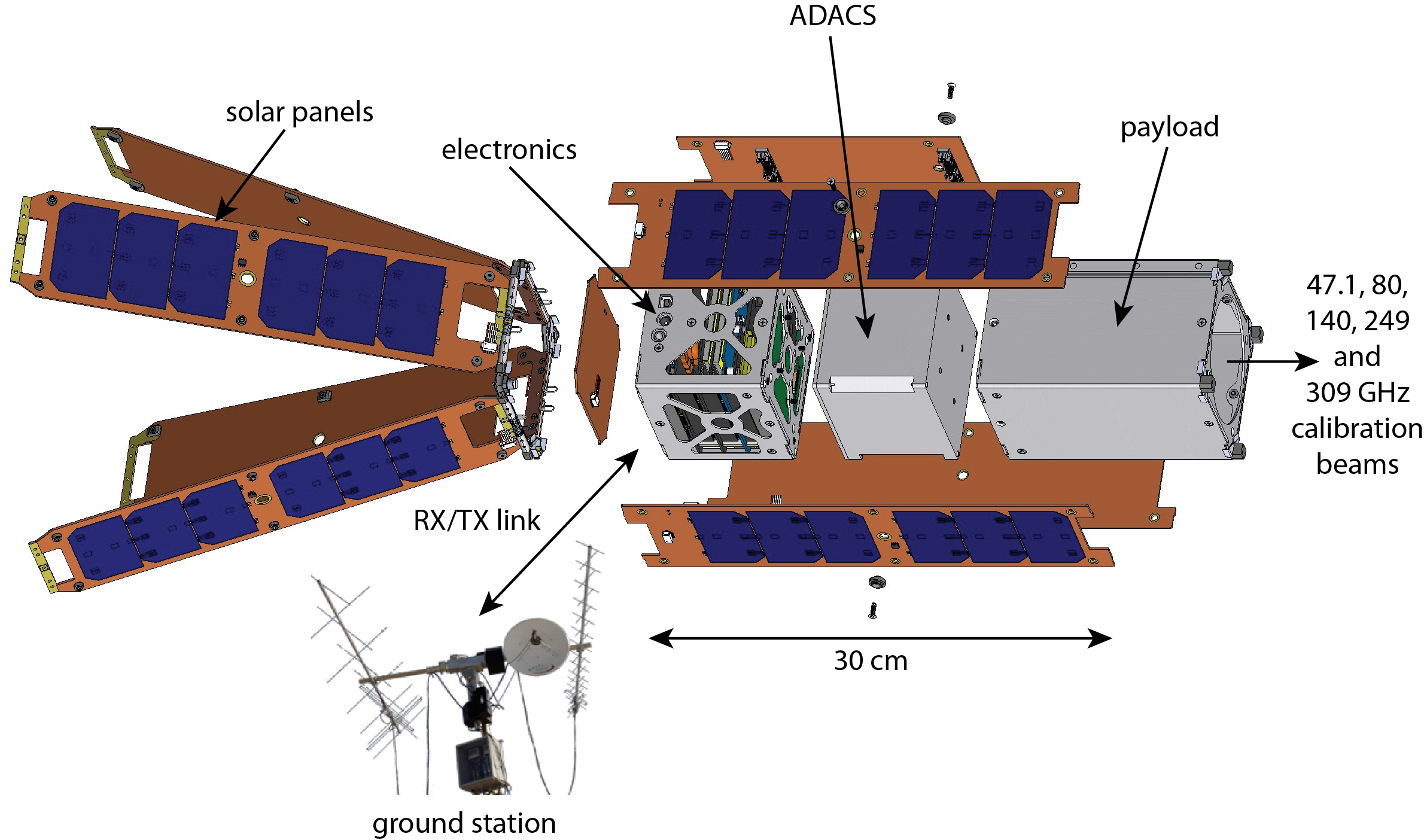}
\caption{
An exploded view of CalSat.
The four CalSat subsystems can all be seen.
The payload section contains the millimeter-wave calibration sources
(see
Figures~\ref{fig:payload_adacs}~\&~\ref{fig:calsat_payload_schematic}).
The attitude determination and control system (ADACS) measures the
CubeSat attitude and rotates the spacecraft as needed (see
Figure~\ref{fig:payload_adacs}).
The electronics subsystem contains the on-board computer and the power
system.
The solar panels are electrically connected to this subsystem.
Note that the power budget was computed using double-sided deployable
solar panels, which are not shown for clarity.
The ground station is used to upload commands and download attitude
and housekeeping data.
The technical details of the hardware components shown in this figure
are given in Section~\ref{sec:calsat_description}.
}
\label{fig:calsat_exploded}
\end{figure}

%%%%%%%%%%%%%%%%%%%%%%%%%%%%%%%%%%%%%%%%%%%%%%%%%%%%%%%%%%%%%%%%%%%%%%%%%%%%

\section{Technical Approach}
\label{sec:technical_approach}

%%%%%%%%%%%%%%%%%%%%%%%%%%%%%%%%%%%%%%%%%%%%%%%%%%%%%%%%%%%%%%%%%%%%%%%%%%%%

\subsection{What are CubeSats?}
\label{sec:cubesats}

%%%%%%%%%%%%%%%%%%%%%%%%%%%%%%%%%%%%%%%%%%%%%%%%%%%%%%%%%%%%%%%%%%%%%%%%%%%%

A CubeSat-based instrument like the one we designed is composed of a
``bus,'' a payload and a ground station.
These components can be seen in Figure~\ref{fig:calsat_exploded}.
The bus includes the mechanical frame, the power system, solar panels,
the on-board computer and electronics, an attitude determination and
control system (ADACS), and a transmitter/receiver.
The payload is the scientific experiment, which is mounted inside the
bus.
For CalSat the payload is the calibration instrument.
The ground station includes a transmitter/receiver and computers, and
it is used to download data and upload commands.
It serves as the point of contact between the research team and the
orbiting spacecraft.

CubeSats essentially get a free ride into orbit as auxiliary payloads.
A Poly Picosatellite Orbital Deployer (P-POD) is installed in a rocket
alongside its primary payload.
CubeSats are loaded into the P-POD, and the P-POD jettisons the
CubeSats at the appropriate time after launch.
The P-POD defines the CubeSat standard, and CubeSats must conform to a
strict list of requirements\footnote{CubeSat requirements document:
  http://www.cubesat.org/index.php/documents/developers} so that they
are compatible with the P-POD.
Among these requirements is a volume requirement.
The fundamental CubeSat size is 10~cm~$\times$~10~cm~$\times$~10~cm,
and this size is referred to as ``1U,'' but larger 2U
(10~cm~$\times$~10~cm~$\times$~20~cm) and 3U
(10~cm~$\times$~10~cm~$\times$~30~cm) sizes also can fit into the
P-POD.

%%%%%%%%%%%%%%%%%%%%%%%%%%%%%%%%%%%%%%%%%%%%%%%%%%%%%%%%%%%%%%%%%%%%%%%%%%%%

\begin{table}[t]
\small
\centering
\begin{tabular}{lcc}
Characteristic                               & Value                        \\
\hline
source frequencies [GHz]                     & 47.1, 80.0, 140, 249 and 309 \\
source spectral width [MHz]                  & $< 1$                        \\
output millimeter-wave power [mW]            & 50, 40, 3.2, 1.2 and 0.45    \\
polarization                                 & linear                       \\
cross-polarization level [dB]                & -60                          \\
horn type                                    & conical                      \\
input waveguide on horn                      & rectangular, single-moded    \\
horn gain [dBi]                              & approximately 20             \\
ADACS steering uncertainty [deg]             & $\sim$1                      \\
ADACS attitude measurement uncertainty [deg] & 0.05                         \\
polarization orientation uncertainty [deg]   & 0.05                         \\
estimated payload mass [kg]                  & 0.8                          \\
estimated total CubeSat mass [kg]            & 3.8                          \\
calculated operating temperature [$^{\circ}$C] & 10 (night) to 30 (day)       \\
orbit altitude [km]                          & 500                          \\
orbital period [hours]                       & 1.6                          \\
orbits per day                               & 14.2                         \\
\hline
\end{tabular}
\caption{CalSat characteristics.  See
  Sections~\ref{sec:calsat_description}~\&~\ref{sec:launch_and_orbit}
  for more detail.}
\label{table:calsat_characteristics}
\end{table}

%%%%%%%%%%%%%%%%%%%%%%%%%%%%%%%%%%%%%%%%%%%%%%%%%%%%%%%%%%%%%%%%%%%%%%%%%%%%

\subsection{CalSat Description}
\label{sec:calsat_description}

%%%%%%%%%%%%%%%%%%%%%%%%%%%%%%%%%%%%%%%%%%%%%%%%%%%%%%%%%%%%%%%%%%%%%%%%%%%%

CalSat is based on a 3U~CubeSat, and all of the required CubeSat
hardware is commercially available.
The primary CalSat characteristics are summarized in
Table~\ref{table:calsat_characteristics}.

%%%%%%%%%%%%%%%%%%%%%%%%%%%%%%%%%%%%%%%%%%%%%%%%%%%%%%%%%%%%%%%%%%%%%%%%%%%%

\begin{figure}[t]
\centering
\includegraphics[width=0.49\textwidth]{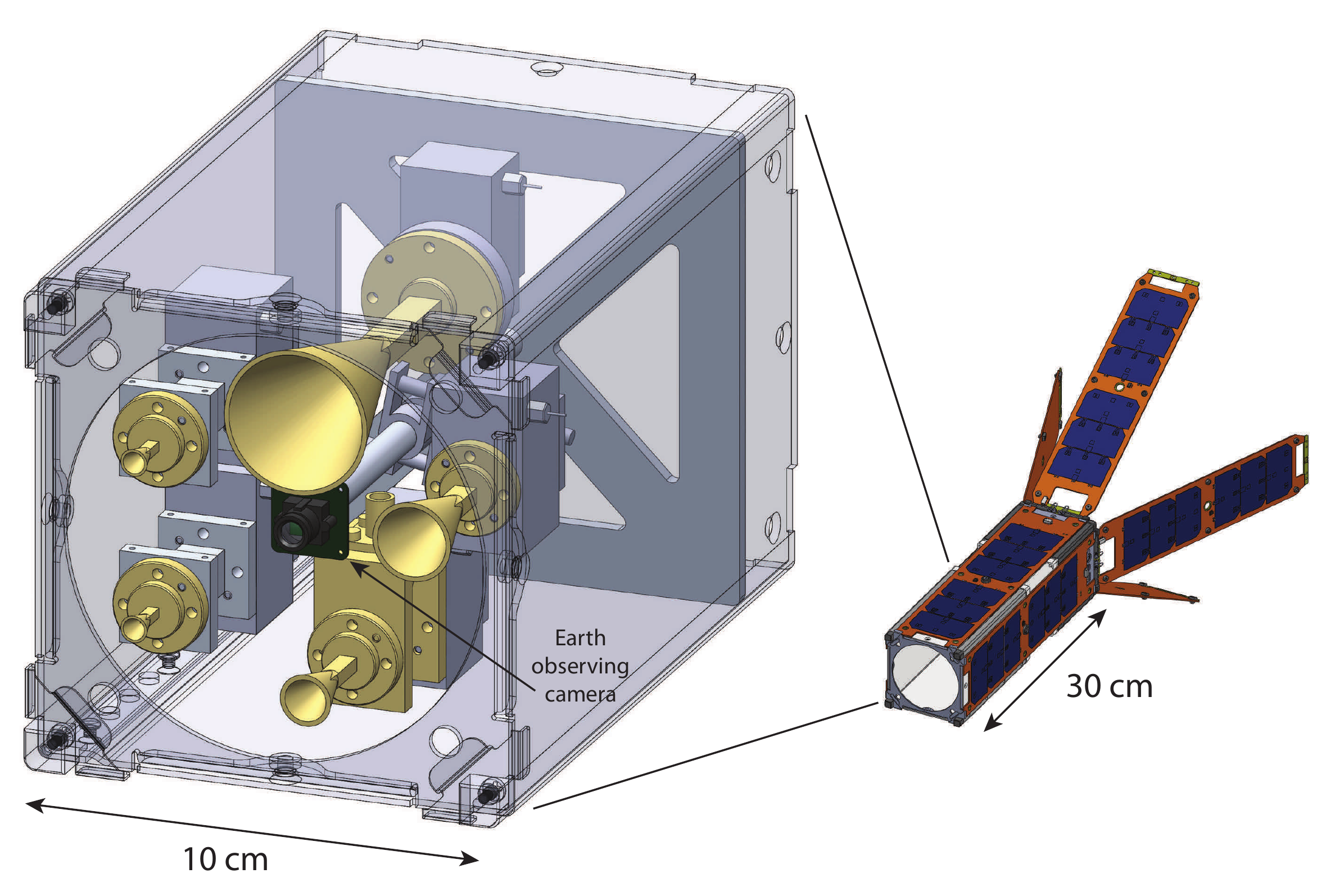}
\includegraphics[width=0.49\textwidth]{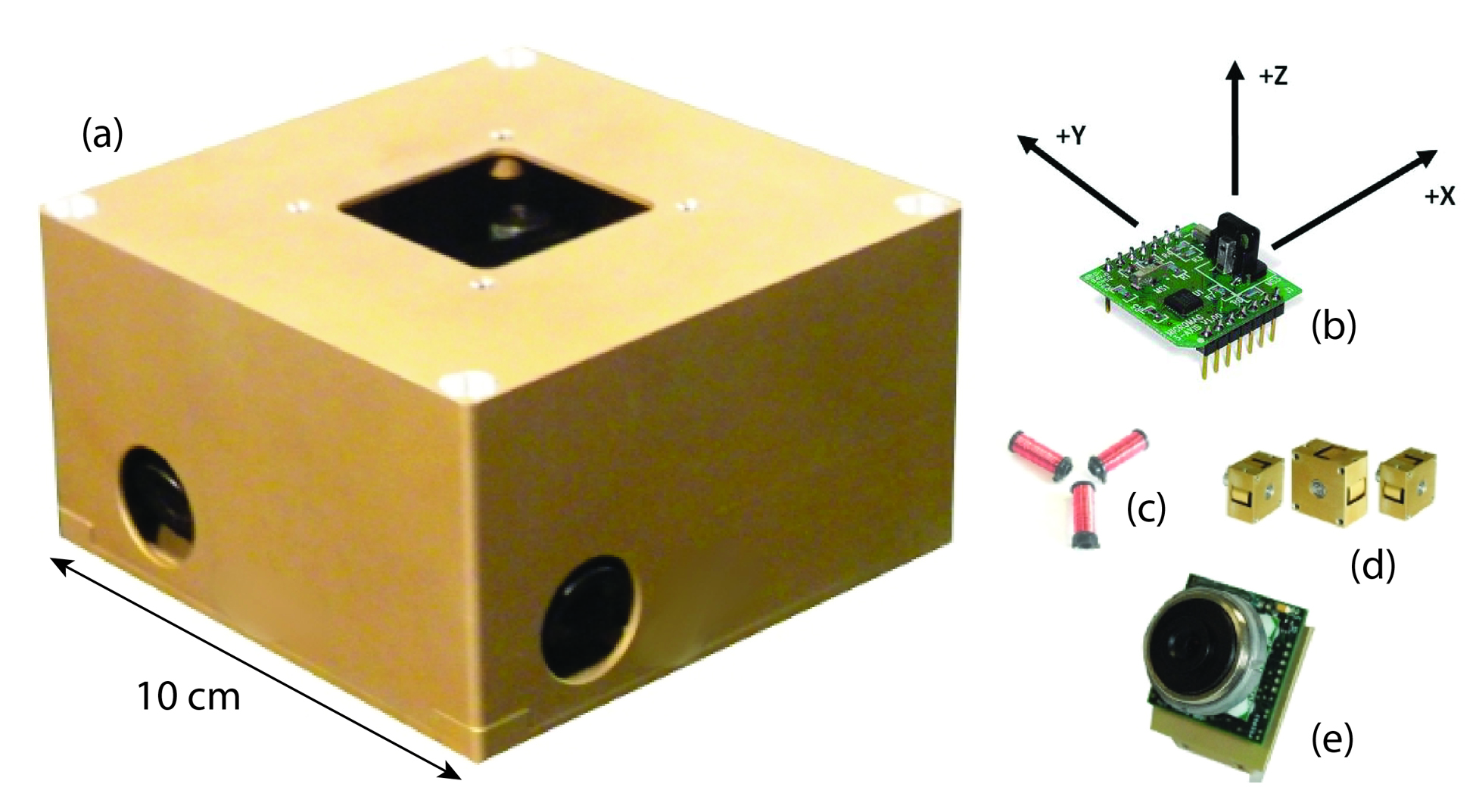}
\caption{
\textbf{Left:} CalSat payload.
The five horn antennas can be seen inside the circular aperture.
From largest to smallest, the horns correspond to 47.1, 80.0, 140, 249
and 309~GHz.
The associated Gunn oscillators and multipliers that produce the
aforementioned five frequencies can be seen at the back of the horns.
A thermal bus connects these sources together and to the payload
walls.
A polarizer mounted at the payload aperture ensures polarization
purity.
A control circuit powers and amplitude modulates the millimeter-wave
sources – each at a slightly different frequency.
The thermal bus, polarizer and control circuit were removed from this
figure to show millimeter-wave sources and to show clearly that they
fit inside the small available volume.
A schematic of the payload is shown in
Figure~\ref{fig:calsat_payload_schematic}.
\textbf{Right:} The attitude determination and control system (ADACS).
CalSat uses the MAI-400SS ADACS from Maryland Aerospace Inc., which is
a turnkey system for CubeSats.
The CubeSat attitude is measured with a magnetometer~(b), star
cameras~(e) and Sun sensors, which are mounted near the solar
panels.
The CubeSat attitude is controlled with three reaction wheels~(d) and
magnetorquer rods~(c), which move the CubeSat by torquing against the
Earth's magnetic field.
}
\label{fig:payload_adacs}
\end{figure}

%%%%%%%%%%%%%%%%%%%%%%%%%%%%%%%%%%%%%%%%%%%%%%%%%%%%%%%%%%%%%%%%%%%%%%%%%%%%

\subsubsection{Bus/ADACS}
\label{sec:bus_adacs}

%%%%%%%%%%%%%%%%%%%%%%%%%%%%%%%%%%%%%%%%%%%%%%%%%%%%%%%%%%%%%%%%%%%%%%%%%%%%

The CalSat design uses the MISC3~3U CubeSat bus from Pumpkin Inc.,
which is a turnkey CubeSat kit that includes all of the aforementioned
bus components.
Pumpkin Incorporated was selected as the bus supplier for CalSat
because their bus meets the pointing performance specification and
more than ten of their CubeSats already have been successfully
launched, therefore their hardware is well tested and comparatively
low risk.
The ADACS, which is manufactured by Maryland Aerospace Incorporated
(MAI), is supplied by Pumpkin Inc.\ as part of their CubeSat kit.
CalSat would use the MAI-400SS, which is a turnkey half-U
(10~cm~$\times$~10~cm~$\times$~5~cm) subsystem that includes three
reaction wheels, a three-axis magnetometer, two star cameras, three
magnetorquer rods, sun sensors and an internal ADACS computer (see
Figure~\ref{fig:payload_adacs}).
The magnetorquer rods are used to desaturate the reaction wheels by
torquing the CubeSat against the magnetic field of the Earth.
To operate this system, the user inputs the desired CalSat quaternion
via the on-board computer, and the ADACS moves CalSat accordingly.
The ADACS also continuously measures and outputs the attitude of
CalSat.
The selected MAI-400SS was designed to track latitude/longitude
positions, so the firmware in this system already includes the
operation mode that CalSat requires.

%%%%%%%%%%%%%%%%%%%%%%%%%%%%%%%%%%%%%%%%%%%%%%%%%%%%%%%%%%%%%%%%%%%%%%%%%%%%

\begin{figure}[t]
\centering
\includegraphics[width=\textwidth]{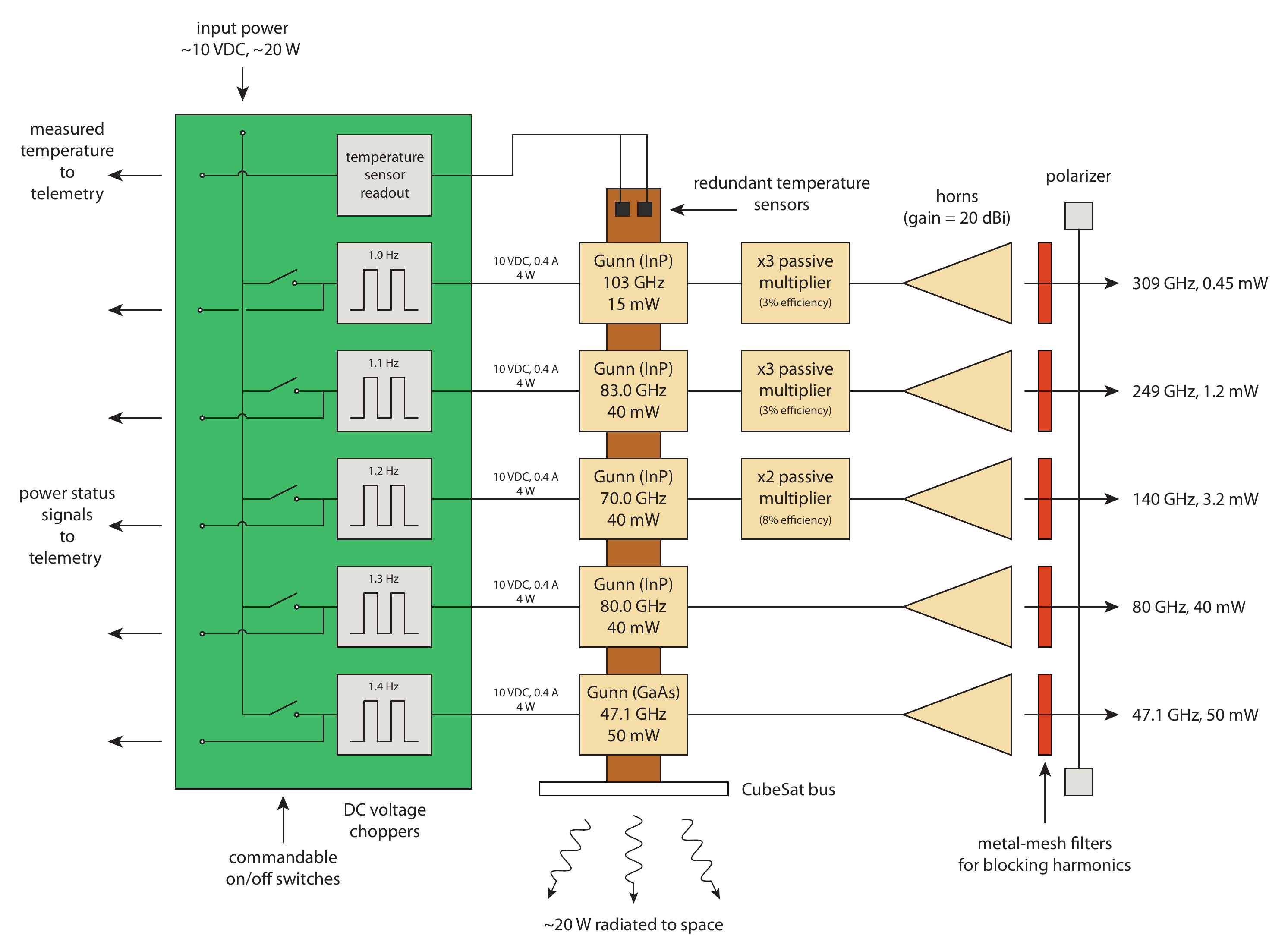}
\caption{
A schematic of the CalSat payload. 
All five millimeter-wave tones are produced with Gunn oscillators.
The three highest-frequency tones are produced with a passive doubler
or tripler.
The millimeter-wave power produced by each chain is given in the
figure label.
All five tones together require only 20~W of input DC power.
}
\label{fig:calsat_payload_schematic}
\end{figure}

%%%%%%%%%%%%%%%%%%%%%%%%%%%%%%%%%%%%%%%%%%%%%%%%%%%%%%%%%%%%%%%%%%%%%%%%%%%%

\subsubsection{Payload}
\label{sec:payload}

%%%%%%%%%%%%%%%%%%%%%%%%%%%%%%%%%%%%%%%%%%%%%%%%%%%%%%%%%%%%%%%%%%%%%%%%%%%%

The calibration source in the payload consists of five amplitude
modulated millimeter-wave ``tones,'' with one each at 47.1, 80.0, 140,
249 and 309~GHz.
These five tones are designed to be well matched to (i) the
observation windows in the atmospheric transmittance spectra, (ii) the
Amateur Satellite Service bands in the Table of Frequency Allocations
used by the Federal Communications Commission
(FCC)\footnote{http://www.fcc.gov} and The National Telecommunications
and Information Administration (NTIA)\footnote{The United States
  Frequency Allocation Chart:
  http://www.ntia.doc.gov/files/ntia/publications/2003-allochrt.pdf},
and (iii) the spectral bands commonly used in CMB polarimeters (more
discussion in Section~\ref{sec:discussion}).
A close-up view of the payload is shown in
Figure~\ref{fig:payload_adacs}.
The 47.1 and 80.0~GHz tones are produced by Gunn diodes directly,
while the 140, 249 and 309~GHz tones are produced by pairing a Gunn
diode with a passive multiplier.
The 140~GHz tone is created by a 70~GHz Gunn diode and a passive
doubler, the 249~GHz tone is created by an 83~GHz Gunn diode and a
passive tripler, and the 309~GHz tone is created by a 103~GHz Gunn
diode and a passive tripler.
Suitable Gunn diodes and multipliers are readily available from Spacek
Labs and Virginia Diodes Inc.~(VDI), respectively.
The tones above 47.1~GHz are produced by InP Gunn diodes because they
output more millimeter-wave power.
This additional power is useful because the required multipliers have
conversion efficiencies of only 3 to 8~\%.
The additional power is not needed at 47.1~GHz, so a
lower-power-output GaAs Gunn diode is used for this tone.
Each tone would have an associated conical horn antenna that would
emit a coherent, linearly polarized beam because the input waveguide
of the horn is rectangular and single-moded.
A small amount of the unwanted cross-polarization, approximately
-30~dB, is produced by the horns adding uncertainty to the orientation
of the calibration source off axis.
Therefore, a wire-grid polarizer\footnote{http://cosmology.ucsd.edu}
is installed at the payload aperture to suppresses this unwanted
cross-polarization by an additional -30~dB or more for all five of the
millimeter-wave sources.
The final cross-polarization level is less than -60~dB, so 99.9999\%
of the power in the CalSat tones is emitted in a single polarization.
The polarizer is installed with its transmission axis parallel to the
co-polarization axis of the horn and slightly tilted to prevent
standing waves between the horn and the polarizer.
The millimeter-wave sources produce harmonics, so metal-mesh low-pass
filters from QMC Instruments are mounted at the horn aperture of each
source to eliminate any unwanted out-of-band radiation.
It is important to emphasize that the CalSat tones are produced by
commercially available components that are based on space-proven
technologies, so the chance of a failure is comparatively low.
Additional technical details are given in
Table~\ref{table:calsat_characteristics} and
Figure~\ref{fig:calsat_payload_schematic}.
Alternative millimeter-wave sources such as broad waveguide-bandwidth
noise sources were initially considered, but they are not compatible
with the Table of Frequency Allocations used by the FCC and NTIA.

%%%%%%%%%%%%%%%%%%%%%%%%%%%%%%%%%%%%%%%%%%%%%%%%%%%%%%%%%%%%%%%%%%%%%%%%%%%%

\subsubsection{Ground Station}
\label{sec:ground_station}

%%%%%%%%%%%%%%%%%%%%%%%%%%%%%%%%%%%%%%%%%%%%%%%%%%%%%%%%%%%%%%%%%%%%%%%%%%%%

An S-band or UHF transmitter/receiver is used to communicate between
the ground station and CalSat.
Ground station hardware is commercially available from vendors like
CubeSatShop\footnote{http://www.cubesatshop.com/} or Clyde
Space\footnote{http://www.clyde-space.com/}.
This radio link is used for downloading data and uploading commands.
The higher frequency S-band link provides a faster data rate and it is
only used if more data bandwidth is necessary.
Amateur UHF or S-band radio frequencies are available for CubeSats.
The CalSat ground station includes a web server.
Spacecraft attitude data and other data products and software
libraries are distributed to the scientific community from this
server.

%%%%%%%%%%%%%%%%%%%%%%%%%%%%%%%%%%%%%%%%%%%%%%%%%%%%%%%%%%%%%%%%%%%%%%%%%%%%

\begin{figure}[t]
\centering
\includegraphics[width=\textwidth]{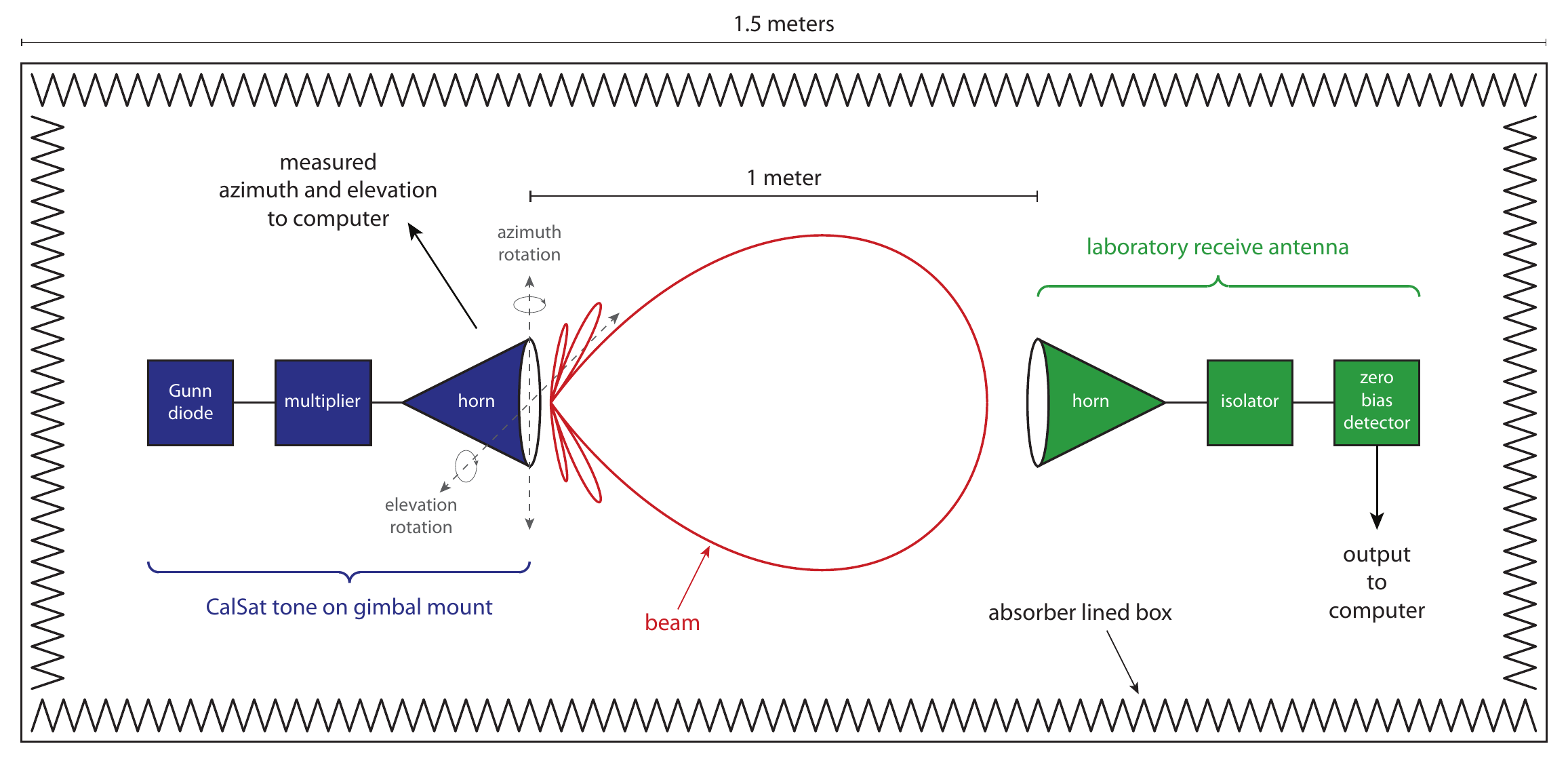}
\caption{Pre-launch beam map characterization measurement.}
\label{fig:test_setup}
\end{figure}

%%%%%%%%%%%%%%%%%%%%%%%%%%%%%%%%%%%%%%%%%%%%%%%%%%%%%%%%%%%%%%%%%%%%%%%%%%%%

\subsection{Pre-launch Testing}
\label{sec:pre-launch_testing}

%%%%%%%%%%%%%%%%%%%%%%%%%%%%%%%%%%%%%%%%%%%%%%%%%%%%%%%%%%%%%%%%%%%%%%%%%%%%

Prior to payload assembly, the horn beams are fully characterized, the
source frequencies measured and the source beams mechanically aligned
with the ADACS in the CalSat bus in the laboratory.
The horn beams are mapped in two dimensions ($\theta,\phi$) with the
antenna pattern measurement system shown in
Figure~\ref{fig:test_setup}.
For this measurement, the CalSat millimeter-wave sources transmit
power and are then pivoted about the phase center of the horn using a
gimbal mount.
A stationary receiver detects the emitted source power at each
orientation.
The gimbal mount is composed of two stepper-motor controlled rotation
stages from Thor Labs.
Both the co-pol and cross-pol beams are measured with this apparatus.
The receiver in this system consists of a horn from Custom Microwave,
a zero-bias Schottky diode detector from VDI and an isolator from
Microwave Resources.
The isolator inserted between the horn and the detector suppresses
standing waves in the system.
The beam mapping apparatus would need to be enclosed in a box lined
with Eccosorb, which is a material that efficiently absorbs
millimeter-waves~\cite{peterson_richards_1984,halpern_1986}.
This baffle would ensure that reflected power does not produce
spurious features in the beam map.

The millimeter-wave source frequencies are measured with a Fourier
transform spectrometer (FTS).
The receivers used in the beam mapping system are reused in this FTS
measurement.
This waveguide-bandwidth FTS measurement yields the precise frequency
of each of the five tones with a very high signal-to-noise ratio.
A second measurement is also done using the CalSat sources and a
broadband 4~K bolometric receiver that can detect radiation below
650~GHz.
For this second measurement, the source power is attenuated somewhat
to match the dynamic range of the detector.
This second measurement reveals any unwanted harmonics in the five
tones, and it also shows that these harmonics are eliminated when the
QMC Instruments filters are added.

The ADACS in the CalSat bus is designed to measure its own orientation
to 0.05~deg.
To ensure that the uncertainty in the polarization orientation of the
CalSat tones is dominated by this ADACS pointing error the
transmission axis of the polarizer in the payload must either be
aligned to the ADACS reference system to better than 0.05~deg or the
systematic offset angle between the two elements must be measured to
better than 0.05~deg.
Calculations show that aligning the polarizer to better than 0.05~deg
would require micron precision, which should be achievable using
standard precision metrology tools.

After CalSat is assembled, and the payload has been shown to work, the
following required tests must be completed before launch: random
vibration, sinusoidal vibration, shock, thermal vacuum cycle, thermal
vacuum bake out, and a hardware configuration test.
These tests are part of the CubeSat Program requirements and are
defined by NASA's Launch Services Program (LSP).
A specialized spacecraft testing facility, such as the facility at Cal
Poly, is required for these tests.

%%%%%%%%%%%%%%%%%%%%%%%%%%%%%%%%%%%%%%%%%%%%%%%%%%%%%%%%%%%%%%%%%%%%%%%%%%%%

\begin{figure}[t]
\centering
$
\begin{array}{cc}
\includegraphics[height=2.2in]{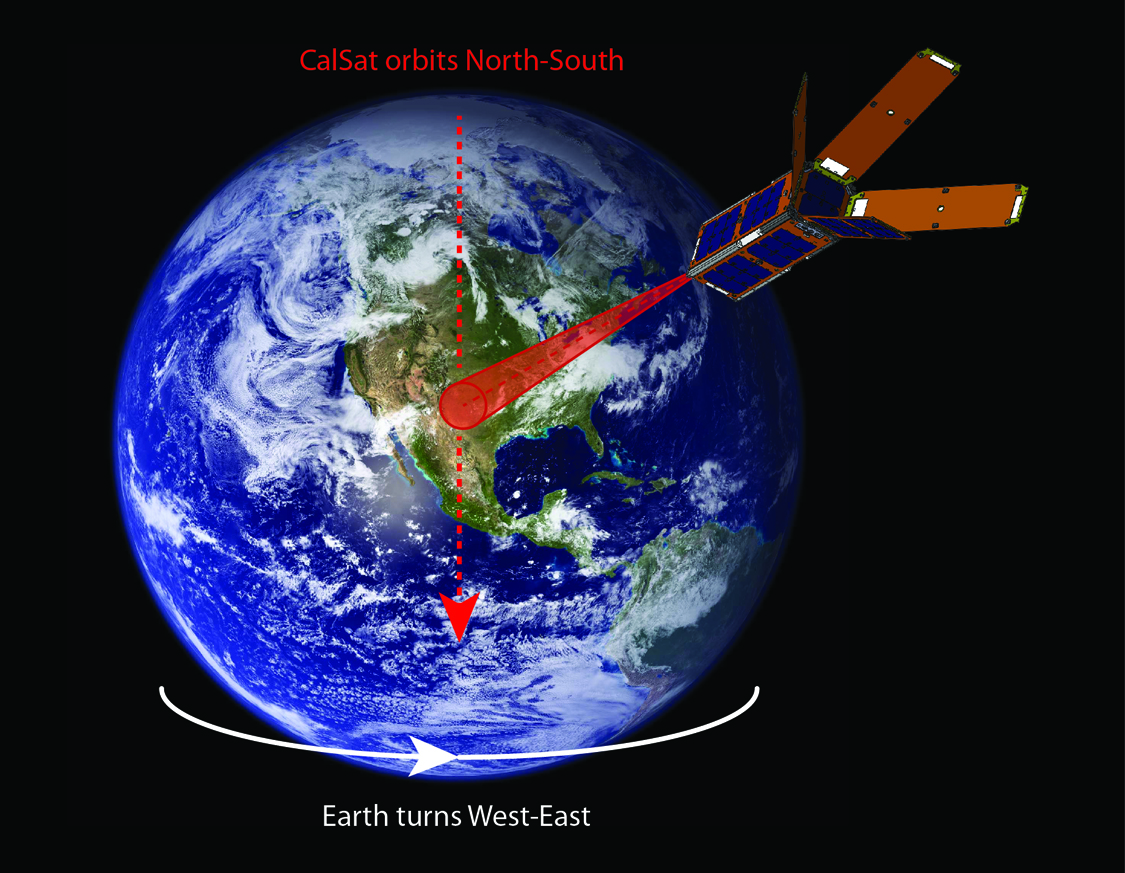} &
\includegraphics[height=2.2in]{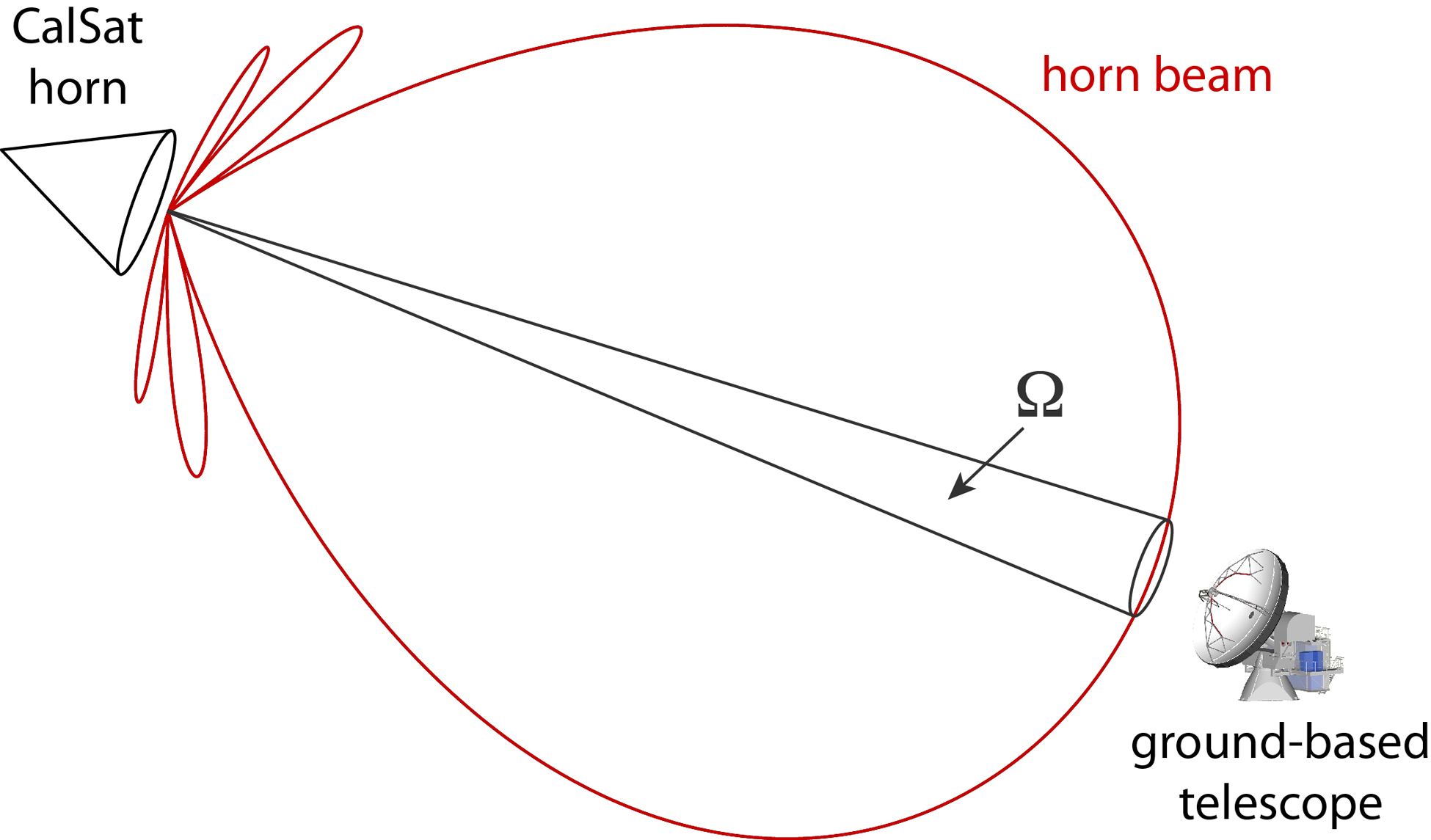} \\
\end{array}
$
\caption{
\textbf{Left:} Schematic of CalSat operation.
For clarity, only one of the five beams is shown.
The calibration beam is pointed at the Earth, while the spacecraft is
in a low-Earth, polar orbit traveling north-south.
The Earth moves west-east below CalSat.
If the precession rate of the orbit is phased appropriately, a
satellite in a polar orbit like this one will, over time, pass over
the entire surface of the Earth (see Figure~\ref{fig:orbits}).
\textbf{Right:} A cross-sectional schematic diagram showing the
coupling between one CalSat horn beam and a ground-based telescope.
Only the emitted power inside the solid angle $\Omega$ is detected.
See Section~\ref{sec:detectability} and Equation~\ref{eq:omega} for
more detail regarding the forecasted CalSat brightness from the power
inside this solid angle.
}
\label{fig:operation}
\end{figure}

%%%%%%%%%%%%%%%%%%%%%%%%%%%%%%%%%%%%%%%%%%%%%%%%%%%%%%%%%%%%%%%%%%%%%%%%%%%%

\subsection{Launch, Orbit and Operations}
\label{sec:launch_and_orbit}

%%%%%%%%%%%%%%%%%%%%%%%%%%%%%%%%%%%%%%%%%%%%%%%%%%%%%%%%%%%%%%%%%%%%%%%%%%%%

CalSat ideally will be placed in a polar orbit.
A polar orbit is a particular type of low-Earth orbit (LEO) where a
spacecraft travels in a north-south direction passing over both the
north and south poles.
Polar orbits are commonly used by Earth-observing satellites.
While a spacecraft orbits in the north-south direction, the Earth
moves beneath it in a west-east direction.
If the precession rate of the orbit is phased appropriately, a
satellite in a polar orbit will, over time, pass over the entire
surface of the Earth.
A conceptual diagram of this kind of orbit is shown in
Figure~\ref{fig:operation}.
A polar orbit for CalSat ensures visibility from observatories in both
the Northern Hemisphere, such as Mauna~Kea in Hawaii and Summit
Station in Greenland~\cite{asada_2014}, and the Southern Hemisphere,
such as the Atacama Desert in Chile, and the South Pole.
CalSat also will be observable by balloon-borne instruments flying
from a range of locations, such as Ft.\ Sumner, New Mexico, Alice
Springs, Australia and McMurdo Station in Antarctica.
This global visibility makes CalSat the only source that can be
observed by all terrestrial and sub-orbital millimeter-wave
observatories.
It is important to emphasize that a LEO does not have fixed Ra/dec
coordinates on the sky, so CalSat would not permanently prevent a
portion of the sky from being observed.

\begin{figure}
\centering
\includegraphics[width=\textwidth]{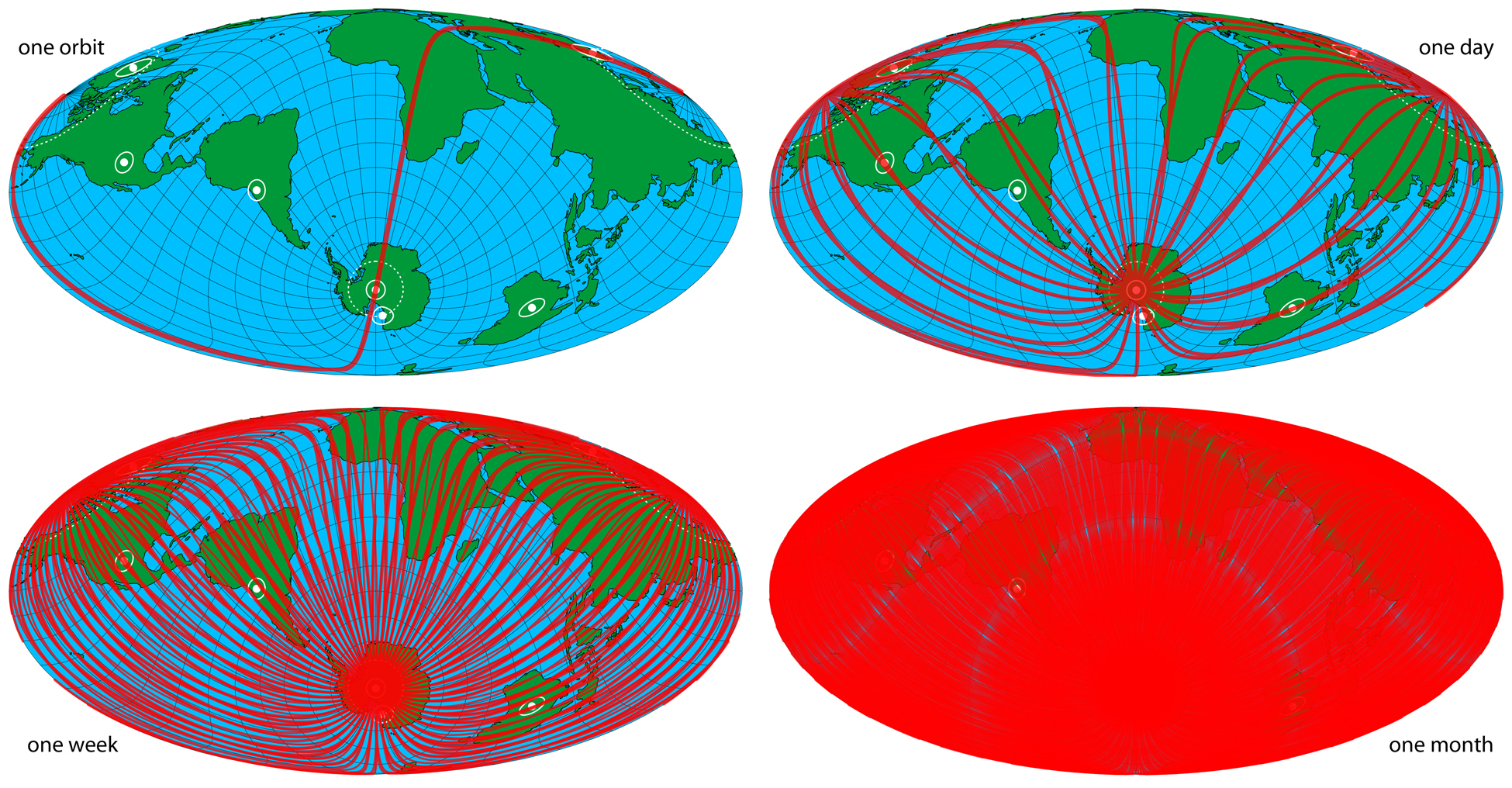}
\caption{
CalSat orbits for 1.6 hours, one day, one week and one month.
Observatories are marked with a white dot.
The radius of the circle surrounding each observatory is 500~km, so
when the red orbit curve is inside the white circle, then the local
zenith angle of CalSat is 45~deg or less at that observatory because
the altitude is also 500~km.
The white dotted lines mark the approximate circumpolar trajectories
of balloons launched from McMurdo Station in Antarctica and Kiruna,
Sweden.
}
\label{fig:orbits}
\end{figure}

\begin{figure}
\centering
\includegraphics[width=0.9\textwidth]{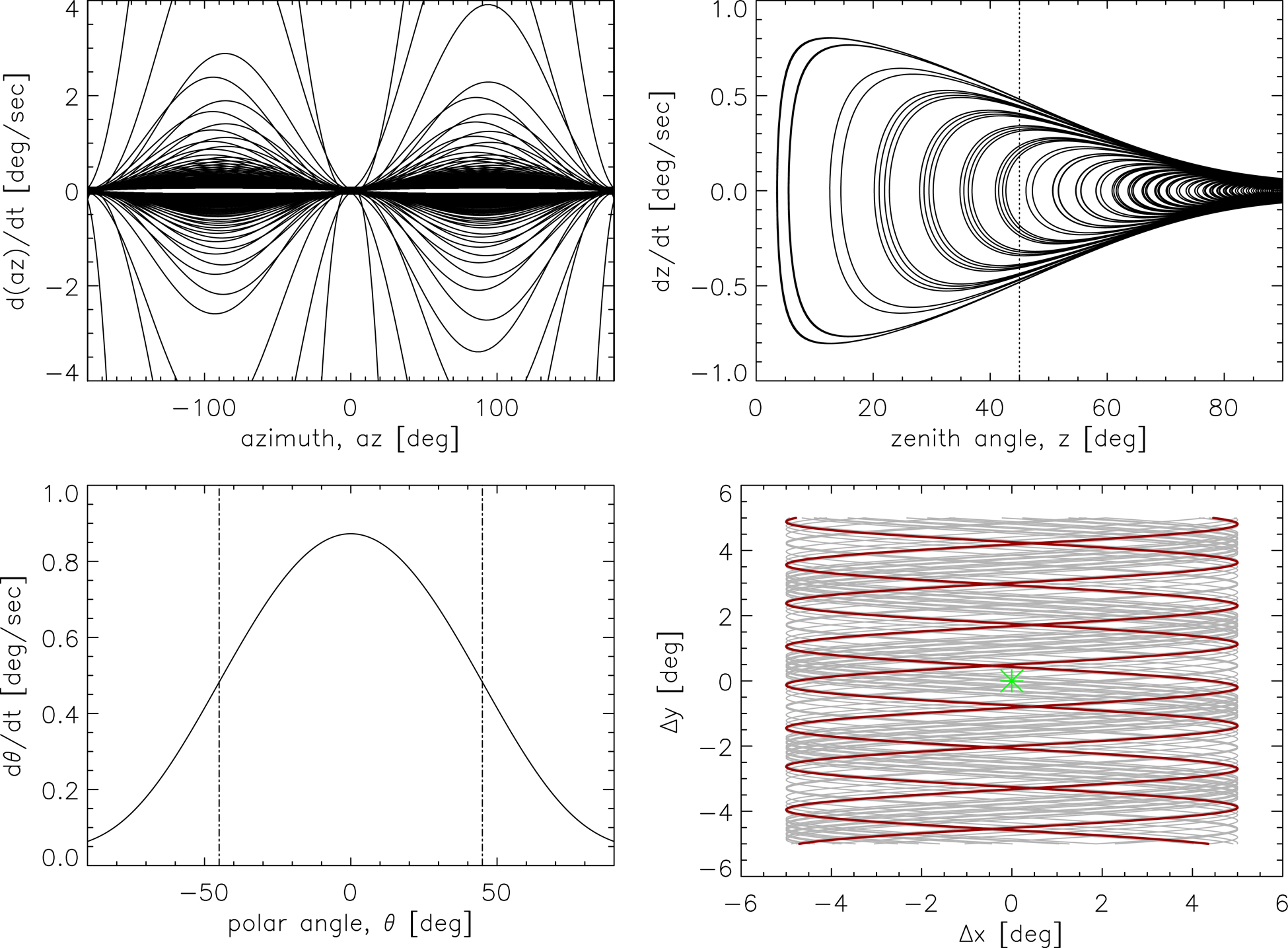}
\caption{
\textbf{Top:} Angular speed of CalSat as observed from the Atacama
Desert in Chile over one week.
The motion is broken up into local azimuth (left) and zenith angle
(right).
\textbf{Bottom Left:} Angular speed of Calsat as observed from the
South Pole.
The azimuth angle is constant for each orbit, so the only motion to
consider is the polar angle $\theta$, which is positive when CalSat is
approaching, zero during transit, and negative when leaving.
\textbf{Bottom Right:} Boresite pointing for one day of CalSat
observations from the South Pole.
Here the telescope tracks CalSat in elevation from 30 to 70~deg, while
scanning sinusoidally in azimuth.
To achieve this pattern in the CalSat-centered frame, both the width of
the azimuth throw of the telescope and the azimuth scan speed increase
as the elevation angle increases, and the elevation angle speed of the
telescope is slightly slower than that of CalSat, so CalSat passed
through the field of view of the instrument.
The red curve shows the scan pattern for a single orbit.
}
\label{fig:angular_speed}
\end{figure}

CalSat would be launched as an auxiliary payload on a DOD, NASA, or
commercial rocket via launch programs such as NASA's Educational
Launch of Nanosatellites
(ELaNa)\footnote{http://www.nasa.gov/offices/education/centers/kennedy/technology/elana\_feature.html},
NASA's CubeSat Launch Initiative
(CSLI)\footnote{http://www.nasa.gov/directorates/heo/home/CubeSats\_initiative.html}
or the Air Force's University Nanosat Program
(UNP)\footnote{http://prs.afrl.kirtland.af.mil/UNP/}.
Because CalSat is launched as a flight of opportunity, the precise
characteristics of the polar orbit, such as the altitude, can not be
specified precisely before the launch opportunity is determined.
Nevertheless, a target orbit is needed so the mission can be designed.
We selected 500~km as the target altitude for the following three
reasons.
First, a 500~km orbit is achievable and viable.
At approximately 1000~km the radiation environment substantially
changes.
Above 1000~km the Van~Allen belts create a radiation environment that
is too hostile for CubeSats.
Below 1000~km the trace amount of atmosphere removes charged particles
making the radiation density suitably low~\cite{smad_1999}.
Several existing CubeSats are in orbits with altitudes between 500 and
800~km.
For comparison, the Hubble Space Telescope orbits at an altitude of
569~km.
Second, at this altitude, air-drag becomes appreciable, and ultimately
it can be used to de-orbit the CubeSat.
Using the Drag Temperature Model (DTM)~\cite{king-hele_1987} we
compute that the maximum expected lifetime of the CalSat orbit is
approximately 16 years.
This lifetime is compatible with the desired CalSat program, and it
complies with the space debris guidelines established by the
Inter-Agency Space Debris Coordination Committee (IADC) and the Office
for Outer Space Affairs, which states that any CubeSat must de-orbit
within 25 years of the end of its mission.
Finally, performance forecasting studies show a 500~km altitude works
well in terms of CalSat detectability (see
Section~\ref{sec:detectability}).
An altitude of 500~km corresponds to an orbital period of 1.6~hours,
which means CalSat passes across the sky at the South Pole 14.2 times
per day, and passes by other observatories several times per week (see
Figure~\ref{fig:orbits}).
The angular speed of CalSat varies with time as it passes by a given
observatory.
In Figure~\ref{fig:angular_speed} we show the angular speed of CalSat
as it passes by observatories at the South Pole and in the Atacama
Desert in Chile.
A typical CMB telescope should be able to track a source moving at
these angular speeds.

%%%%%%%%%%%%%%%%%%%%%%%%%%%%%%%%%%%%%%%%%%%%%%%%%%%%%%%%%%%%%%%%%%%%%%%%%%%%

%%%%%%%%%%%%%%%%%%%%%%%%%%%%%%%%%%%%%%%%%%%%%%%%%%%%%%%%%%%%%%%%%%%%%%%%%%%%

%%%%%%%%%%%%%%%%%%%%%%%%%%%%%%%%%%%%%%%%%%%%%%%%%%%%%%%%%%%%%%%%%%%%%%%%%%%%

The location of CalSat is tracked by the North American Aerospace
Defense Command (NORAD) and quantified by a six element state vector
that includes position and velocity information.
NORAD would provide the state vector, which is uploaded to CalSat by
the ground station.
The ADACS would then use this state vector in conjunction with sensor
information to determine the attitude of the spacecraft.
The on-board computer would also take this position information,
determine which observatory is closest, and then point the calibration
beams at that observatory.
During operation, the CalSat beams are pointed at CMB observatories
using the ADACS and then modulated at approximately 1~Hz.
By modulating the amplitude of the tones, it is easier to detect them
if the beam-filling atmospheric emission is fluctuating.
The calibration beam attitude is updated in real time as CalSat moves
past the observatory to ensure the millimeter-wave source beams are
fixed on that observatory.
It is important to point out that the polarization orientation of the
millimeter-wave sources is not fixed in Ra/dec, and it will slowly
drift relative to the reference frame of the observer as CalSat moves
across the sky.
However, this polarization orientation drift is always measured to
0.05~deg using the on-board star cameras, so this effect can be
precisely accounted for during data analysis in the same way
orientation drift is treated with celestial sources.

The measured attitude is recorded as a function of time and this
time-ordered data is telemetered to the ground station each orbit.
The downloaded data would immediately be made publicly available via a
web server.
The measured source beam patterns (see
Section~\ref{sec:pre-launch_testing}) and CalSat attitude data would
then be used by the CMB community in the analysis of their calibration
measurements.
End users of our data would need the azimuth, elevation, polarization
orientation and the anticipated source brightness of CalSat as a
function of time.
To facilitate use of our data, the CalSat team would write a software
library that would compute these needed quantities for any given
latitude and longitude, and this software library is distributed along
with the data.

%%%%%%%%%%%%%%%%%%%%%%%%%%%%%%%%%%%%%%%%%%%%%%%%%%%%%%%%%%%%%%%%%%%%%%

\begin{table}[t]\footnotesize
\small
\centering
\begin{tabular}{lccc}
                  & peak power [W] & duty cycle [\%] & average power [W] \\
\hline
payload           & 20             & 20              & 4.0               \\
ADACS (MAI-400SS) & 2.2            & 100             & 2.2               \\
transceiver (Tx)  & 5.5            & 10              & 0.55              \\
transceiver (Rx)  & 0.25           & 100             & 0.25              \\
on-board computer & 2.0            & 100             & 2.0               \\
\hline
total             & 30             &                 & 9.0                \\
\end{tabular}
\caption{\footnotesize{CalSat power budget.  The orbit average power
    was computed to be 12~W, and this should be compared with the 9~W
    of total average power in this table.  CalSat should have 3.0~W of
    margin.}}
\label{table:power_budget}
\end{table}

%%%%%%%%%%%%%%%%%%%%%%%%%%%%%%%%%%%%%%%%%%%%%%%%%%%%%%%%%%%%%%%%%%%%%%

\subsection{Power Budget}
\label{sec:power_budget}

%%%%%%%%%%%%%%%%%%%%%%%%%%%%%%%%%%%%%%%%%%%%%%%%%%%%%%%%%%%%%%%%%%%%%%%%%%%%

A breakdown of the power budget for CalSat is given in
Table~\ref{table:power_budget}.
For the mission to be sustainable, the orbit average power (OAP) must
be greater than the average power consumption.
In this limit, the solar panels are able to, on average, both power
the CubeSat and keep the on-board battery charged.
Here we computed the OAP as the sun/eclipse ratio times the average
available power from the solar panels during sunlight time, assuming
the Earth's albedo contribution is negligible during eclipse time.
At 500~km, the sun/eclipse ratio is 0.62. 
Given the number of solar panels in the MISC3 CubeSat, the solar panel
configuration (see Figure~\ref{fig:calsat_exploded}), and assuming
each 3U panel can provide 2~W of power, we computed that between 13
and 24~W is available depending on the orbit angle.
The average available power is 20~W during sunlight time, which means
the OAP is 12~W.
Table~\ref{table:power_budget} shows that the average power
consumption is conservatively 9~W, so the CalSat mission is
sustainable with approximately 3~W of margin.

%%%%%%%%%%%%%%%%%%%%%%%%%%%%%%%%%%%%%%%%%%%%%%%%%%%%%%%%%%%%%%%%%%%%%%%%%%%%

\begin{figure}
\centering
$
\begin{array}{c}
\includegraphics[width=0.6\textwidth]{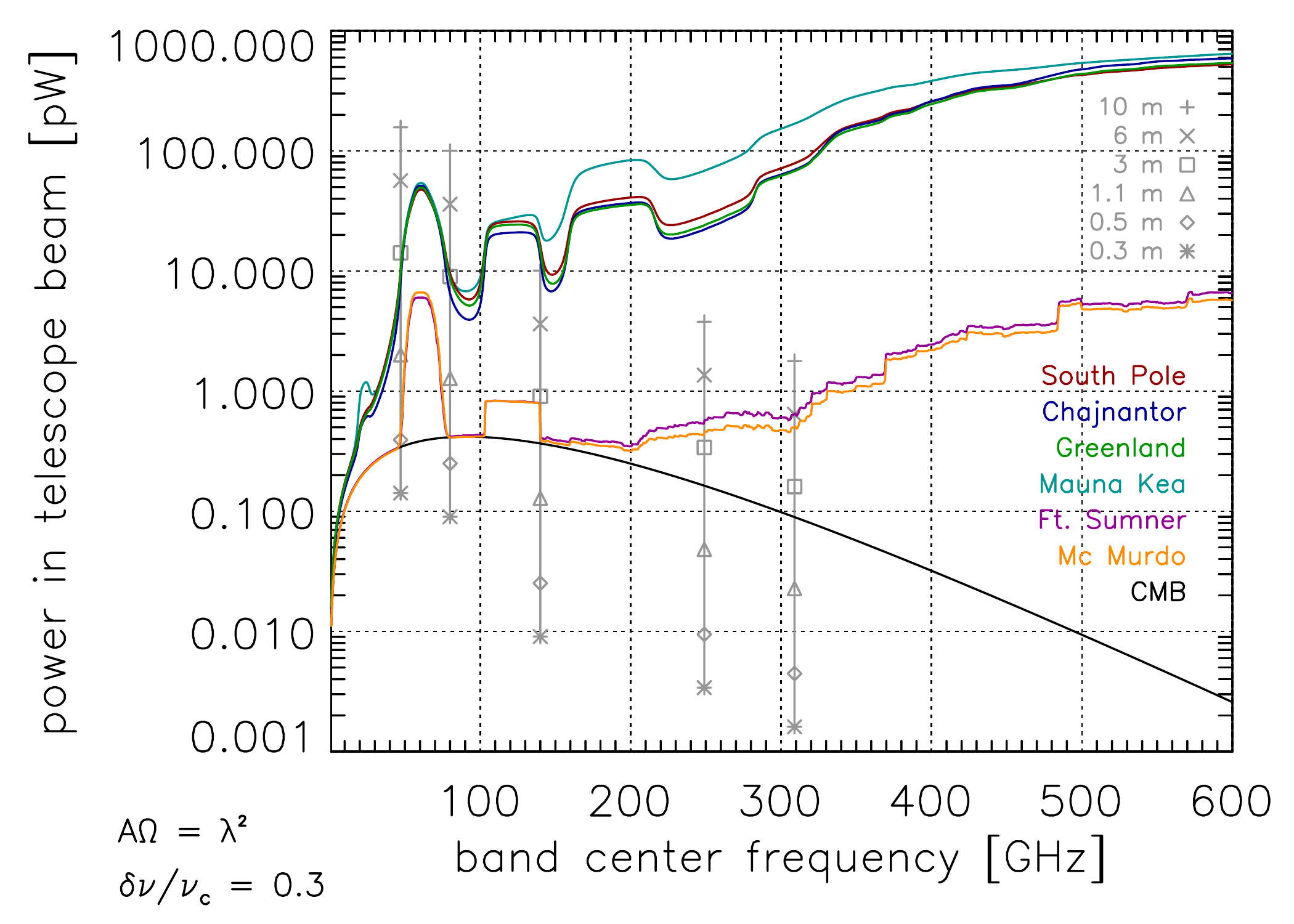} \\
\includegraphics[width=0.6\textwidth]{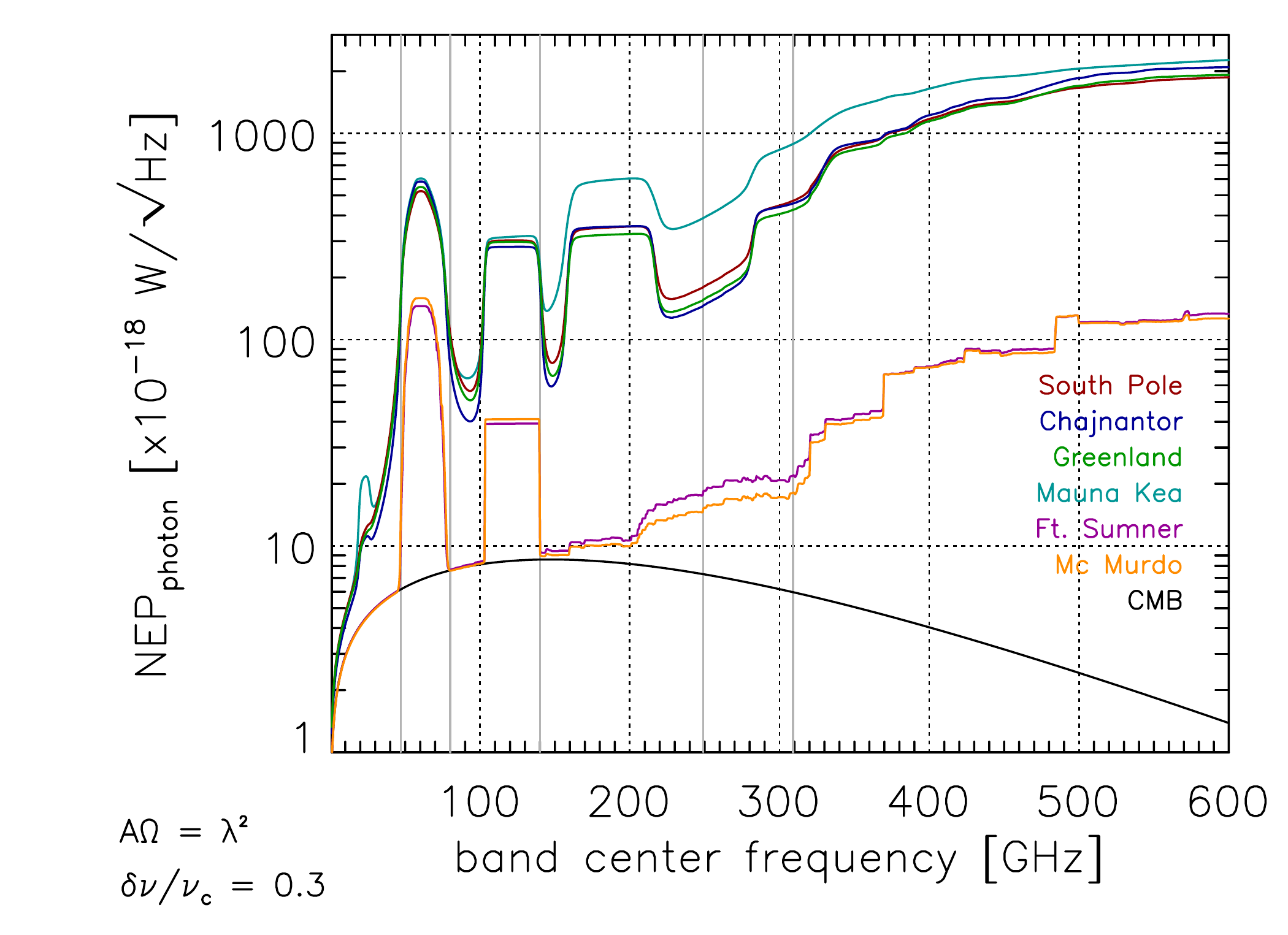} \\
\end{array}
$
\caption{CalSat detectability.
These figures together show (i) that CalSat is detectable with high
signal-to-noise, and (ii) that the power loading from CalSat can be
compared with the power loading from the background, which is ideal.
In the top panel we show the amount of millimeter-wave power CalSat
would deliver to the telescope aperture.
Six different telescope aperture diameters are indicated with symbols.
The curves in this panel show the forecasted level of background power
as a function of spectral band center frequency assuming some typical
instrument characteristics and observatory conditions that are given
in Section~\ref{sec:detectability}.
For this analysis the background power comes from the CMB and the
atmosphere.
The bottom panel shows the expected photon NEP as a function of
spectral band center frequency.
As an example, a 150~GHz polarimeter with a 2.0~m aperture sited at
the Llano de Chajnantor Observatory in the Atacama Desert would
receive 10~pW of power from the sky, nearly 1~pW of power from CalSat,
and the noise rms is $\sim10^{-5}$~pW (assuming one second of
integration with NEP$_{photon} =
10\times10^{-18}$~W/$\sqrt{\mbox{Hz}}$).
Therefore the CalSat signal-to-noise ratio is $10^{5}$ (or 50~dB) for
this configuration.
More signal-to-noise values computed from the information in these
plots are given in Table~\ref{table:signal_to_noise}.}
\label{fig:loading_nep}
\end{figure}

%%%%%%%%%%%%%%%%%%%%%%%%%%%%%%%%%%%%%%%%%%%%%%%%%%%%%%%%%%%%%%%%%%%%%%%%%%%%

\begin{table}[t]
%\small
\footnotesize
\centering
\begin{tabular}{cccccccc}
 Frequency & Aperture & & & & & & \\
{[GHz]} & [m] & South Pole & Chajnantor & Greenland & Mauna Kea & Ft. Sumner & McMurdo \\
\hline
47.1 & 0.3 & 31 & 31 & 31 & 31 & 42 & 41 \\
47.1 & 0.5 & 35 & 36 & 35 & 35 & 47 & 46 \\
47.1 & 1.1 & 42 & 43 & 43 & 42 & 54 & 53 \\
47.1 & 3 & 51 & 51 & 51 & 51 & & \\
47.1 & 6 & 57 & 58 & 57 & 57 & & \\
47.1 & 10 & 61 & 62 & 61 & 61 & & \\
\hline
80.0 & 0.3 & 31 & 32 & 31 & 31 & 42 & 42 \\
80.0 & 0.5 & 35 & 37 & 35 & 35 & 47 & 47 \\
80.0 & 1.1 & 42 & 44 & 43 & 43 & 54 & 54 \\
80.0 & 3 & 51 & 52 & 51 & 51 & & \\
80.0 & 6 & 57 & 58 & 57 & 57 & & \\
80.0 & 10 & 61 & 63 & 61 & 61 & & \\
\hline
140 & 0.3 & 18 & 18 & 18 & 17 & 30 & 30 \\
140 & 0.5 & 22 & 23 & 22 & 22 & 34 & 34 \\
140 & 1.1 & 29 & 30 & 30 & 29 & 41 & 41 \\
140 & 3 & 38 & 38 & 38 & 37 & & \\
140 & 6 & 44 & 44 & 44 & 43 & & \\
140 & 10 & 48 & 49 & 49 & 48 & & \\
\hline
249 & 0.3 & 14 & 15 & 15 & 11 & 24 & 25 \\
249 & 0.5 & 19 & 20 & 19 & 15 & 29 & 29 \\
249 & 1.1 & 26 & 27 & 26 & 22 & 36 & 37 \\
249 & 3 & 34 & 35 & 35 & 31 & & \\
249 & 6 & 40 & 41 & 41 & 37 & & \\
249 & 10 & 45 & 46 & 45 & 41 & & \\
\hline
309 & 0.3 & 7 & 7 & 7 & 4 & 20 & 21 \\
309 & 0.5 & 11 & 11 & 12 & 9 & 25 & 25 \\
309 & 1.1 & 18 & 18 & 19 & 16 & 32 & 32 \\
309 & 3 & 27 & 27 & 27 & 24 & & \\
309 & 6 & 33 & 33 & 33 & 30 & & \\
309 & 10 & 37 & 37 & 38 & 35 & & \\
\hline
\end{tabular}
\caption{
Forecasted CalSat signal-to-noise ratio \textbf{in dB} for various
observatories and telescope apertures.
We assume the telescope zenith angle is 45~deg, and the CalSat
altitude is 500~km.
Here the signal-to-noise ratio is the peak signal divided by standard
error of the mean for one second of integration.
These ratios improve as the integration time increases and the
telescope zenith angle approaches 0~deg.
}
\label{table:signal_to_noise}
\end{table}

%%%%%%%%%%%%%%%%%%%%%%%%%%%%%%%%%%%%%%%%%%%%%%%%%%%%%%%%%%%%%%%%%%%%%%%%%%%%

\begin{figure}[t]
\centering
\includegraphics[width=0.8\textwidth]{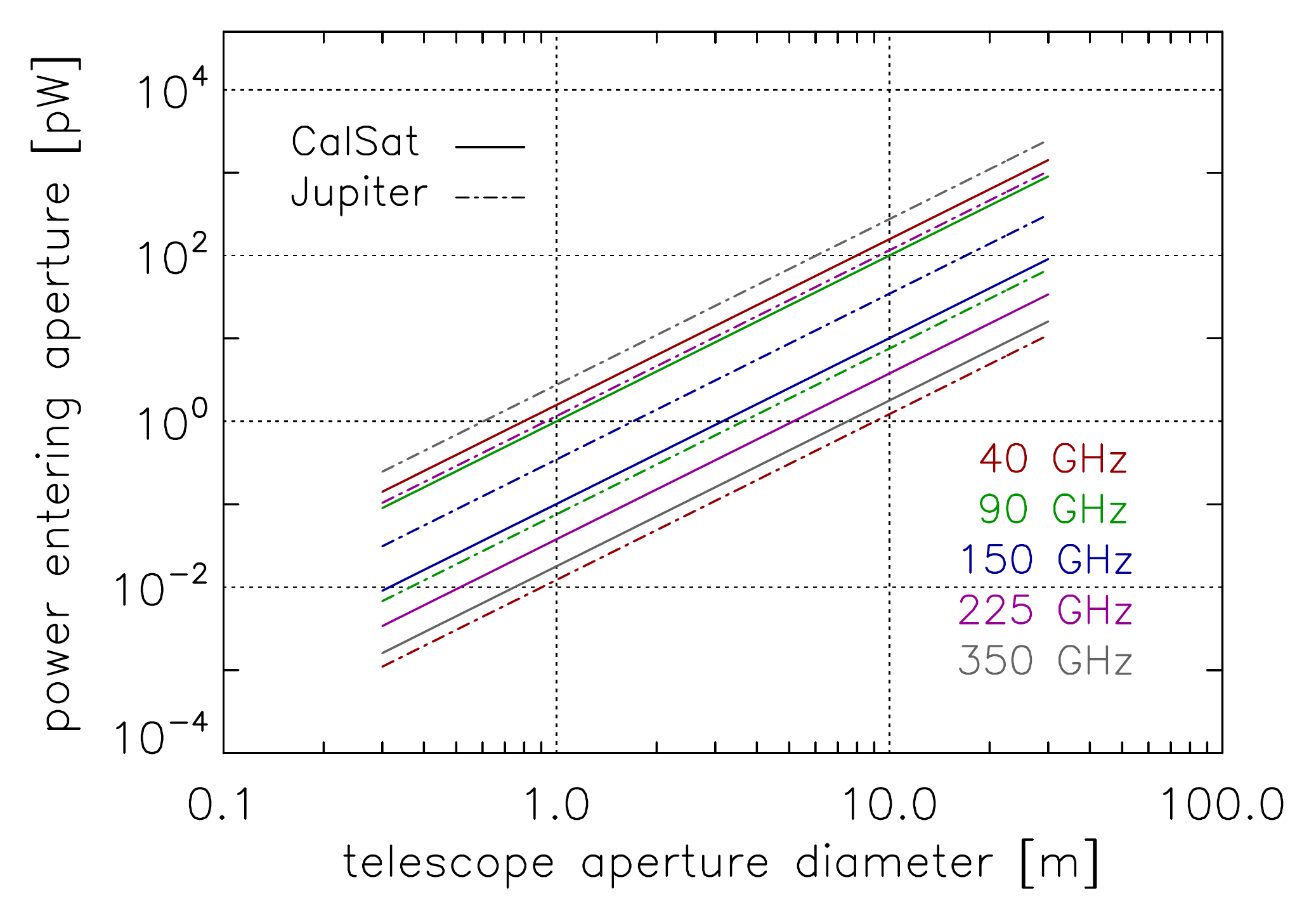}
\caption{ 
Comparison between the power delivered by CalSat and the total power,
integrated across the spectral band, from Jupiter.
This comparison was computed for a range of telescope aperture
diameters.
When computing the curves for Jupiter we assumed A$\Omega =
\lambda^2$, $\delta\nu/\nu_c = 0.3$ and the brightness temperature is
155~K for the 40~GHz spectral band and 173~K for the other spectral
bands~\cite{weiland_2011,griffin_1986}.
The CalSat curves are discussed in Section~\ref{sec:detectability}.
An additional comparison between CalSat and Tau~A is given in
Table~\ref{table:tau_a}.
}
\label{fig:calsat_and_jupiter}
\end{figure}

%%%%%%%%%%%%%%%%%%%%%%%%%%%%%%%%%%%%%%%%%%%%%%%%%%%%%%%%%%%%%%%%%%%%%%%%%%%%

\section{Detectability}
\label{sec:detectability}

%%%%%%%%%%%%%%%%%%%%%%%%%%%%%%%%%%%%%%%%%%%%%%%%%%%%%%%%%%%%%%%%%%%%%%%%%%%%

% What results are expected?  Is CalSat even detectable?

For CalSat to be useful, the detected calibration signal needs to be
large when compared with the noise level of the calibration
measurement.
It is difficult to forecast the precise characteristics of every
instrument that might observe CalSat.
Therefore, to assess the detectability, three primary assumptions were
made.
First, we assumed the aperture diameter of any instrument observing
CalSat is between 0.3 and 10~m, which is the current range for CMB
experiments.
Second, we assumed A$\Omega = \lambda^2$ and $\delta\nu/\nu_c = 0.3$.
Here, $A$ is the aperture area, $\Omega$ is the solid angle of the CMB
telescope beam, and $\nu$ is the frequency of the incoming radiation
with $\nu_c$ being the center frequency of the spectral band of the
polarimeter.
Third, we assumed the noise in the calibration measurement is
dominated by photon noise from the CMB for balloon-borne measurements
and the atmosphere for ground-based measurements.

To estimate the photon noise, spectral radiance curves were computed
using the publicly available \textit{am} atmospheric modeling software
package\footnote{The \textit{am} software:
  https://www.cfa.harvard.edu/$\sim$spaine/am/}.
We used \textit{am} version 8.0, which implements O$_2$ line mixing
and non-resonant absorption near 60~GHz.
Earlier versions of \textit{am} did not model these effects, so these
earlier versions had accuracy issues, which do not affect our
calculations.
The atmospheric profiles for the following six observatories were used
in these simulations: South Pole, Chajnantor, Summit Station in
Greenland, Mauna~Kea, and balloon-borne observatories near
Ft.\ Sumner, New Mexico, and McMurdo in Antarctica.
For all sites, the zenith angle for observations was set to
45~deg.
The altitude of the balloon-borne observatories was assumed to be
30~km.
For the South Pole, Chajnantor, Greenland, and Mauna~Kea
observatories, the ambient temperature and PWV was assumed to be 230,
275, 248 and 275~K, and 1, 1, 1 and 4~mm, respectively.
Given these conservative assumptions, the background power in the
telescope beam and the associated photon noise equivalent power (NEP)
were computed.
The background power in the telescope beam was computed using
\begin{equation}
P_{photon}~=~\int Q(\nu)~d\nu~~[\mbox{W}],~~\mbox{where}~~Q(\nu) = \alpha~{\cal T}(\nu)~A \Omega~{\cal B}(\nu).
\label{eq:power}
\end{equation}
and the photon NEP was computed using
\begin{equation}
NEP_{photon}^2~=~2 \int h \nu~Q(\nu)~d\nu~+~2 \int
\frac{c^2~Q(\nu)^2}{m~A \Omega~\nu^2}~d\nu~~[\mbox{W}^2/\mbox{Hz}].
\label{eq:nep}
\end{equation}
These equations are commonly used and their derivation can be found in
the literature \cite{lamarre_1986,richards_1994}.
For our calculations, ${\cal B}(\nu)$ is the spectral radiance
computed by \textit{am}, ${\cal T}(\nu)$ is the spectral band-pass
filter, which is a top-hat with a width equal to $\delta\nu/\nu_c =
0.3$ and A$\Omega = \lambda^2$.
We assumed observations are made with detectors that are sensitive to
a single-polarization, so $\alpha = 0.5$ and $m = 1$.
The results are shown in Figure~\ref{fig:loading_nep} and
Table~\ref{table:signal_to_noise}.
The expected millimeter-wave power from CalSat at the telescope
aperture was computed using the following equation:
\begin{equation}
P_{CalSat}~=~\frac{P_{Gunn}}{4\pi}\,10^{g/10}\,\Omega~~[\mbox{W}],~~\mbox{where}~~\Omega = \pi \left( \tan^{-1} \left(\frac{D}{2d} \right) \right)^2.
\label{eq:omega}
\end{equation}
Here, $P_{Gunn}$ is the millimeter-wave power from the source, $g$ is
the horn gain in dBi, $D$ is the aperture diameter of the telescope,
and $d$ is the distance between CalSat and the telescope; $\Omega$ is
the solid angle subtended by the telescope aperture as viewed from
CalSat, and it is schematically shown in Figure~\ref{fig:operation}.
The power arriving in the telescope aperture from CalSat is plotted
versus telescope aperture diameter in
Figure~\ref{fig:calsat_and_jupiter}.
For comparison, the power arriving in the telescope aperture from
Jupiter, which is a bright and commonly used {\it unpolarized}
calibrator, was also computed and plotted alongside the CalSat curves
in this Figure.

This analysis shows that (i) the brightness of CalSat is similar to
the brightness of the background, (ii) the brightness of CalSat is
comparable to Jupiter, and (iii) the signal-to-noise ratio for one
second of integration is typically between 1,000 and 100,000.
These are all characteristics of an ideal calibration source.
If CalSat were significantly brighter than the background then the
signal would likely saturate the detectors (assuming TES bolometers
are being used, which is the current standard at these frequencies).

%%%%%%%%%%%%%%%%%%%%%%%%%%%%%%%%%%%%%%%%%%%%%%%%%%%%%%%%%%%%%%%%%%%%%%%%%%%%

\begin{figure}[t]
\centering
\includegraphics[width=0.8\textwidth]{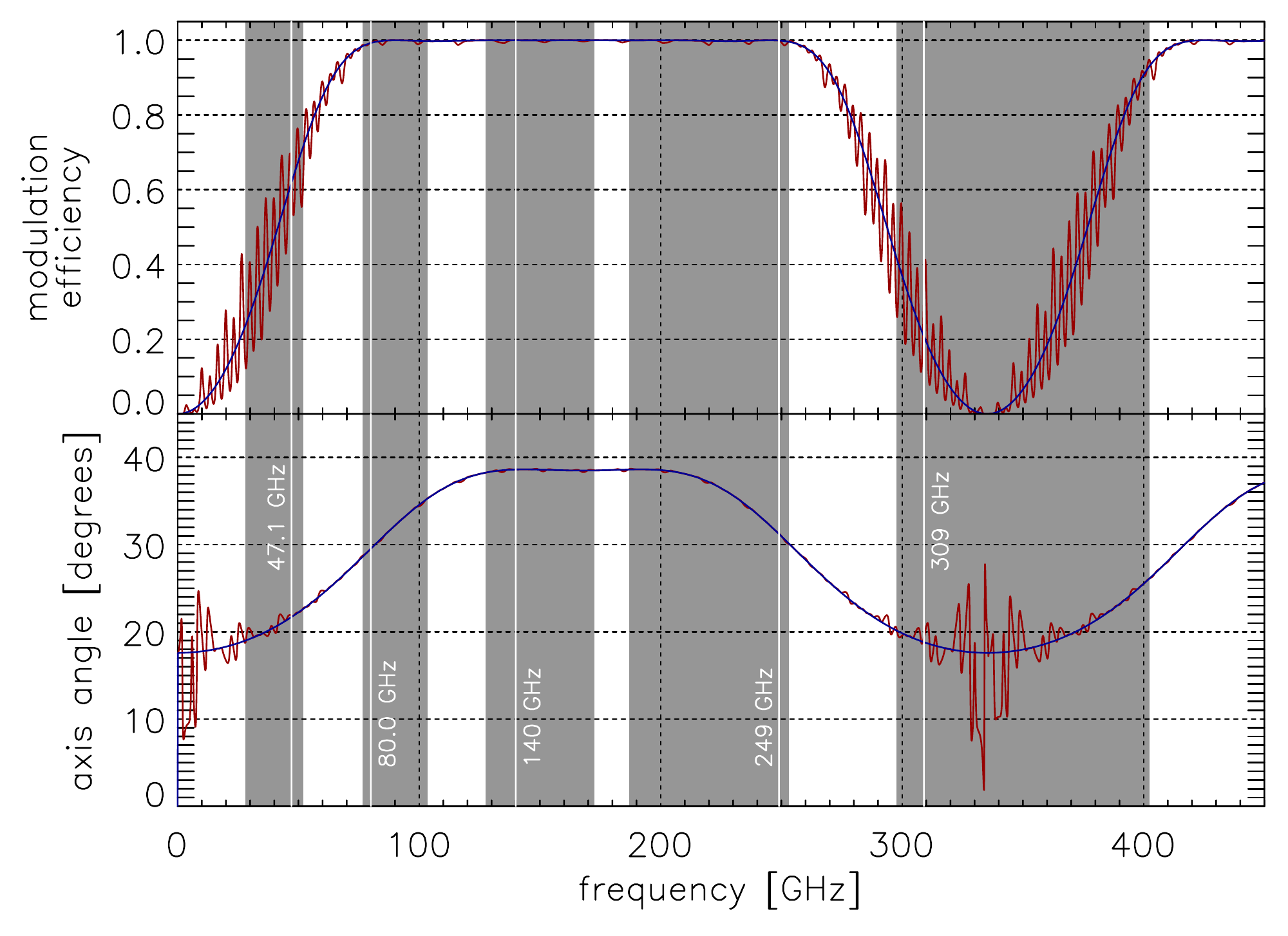}
\caption{
One example showing the utility of CalSat.
The plot shows simulated modulation efficiency and associated
polarimeter axis curves for a five-crystal achromatic half-wave plate
polarimeter~\cite{savini_2006}.
This kind of polarimeter has nearly perfect modulation efficiency
across three spectral bands, which is attractive.
However, the polarimeter axis varies with frequency by approximately
10~deg, and this axis angle needs to be calibrated to better than
0.2~deg.
An observation of CalSat could constrain the axis-angle vs.\ frequency
curve for this instrument.
Note that if a celestial broadband source, such as Tau~A, were used
instead, the frequency spectrum of this celestial source would have to
be precisely measured first to produce the same kind of constraint.}
\label{fig:hwp_axis}
\end{figure}

The CalSat instrument configuration presented in this paper was
designed to support the following ongoing experiments and any
associated follow-up experiments from these collaborations:
ACT~\cite{naess_2014},
BICEP/KECK~\cite{ahmed_2014,buder_2014},
CLASS~\cite{essinger-hileman_2014},
GroundBIRD~\cite{tajiman_2012},
EBEX~\cite{reichborn-kjennerud_2010},
PIPER~\cite{lazear_2014},
POLARBEAR~\cite{arnold_2014,tomaru_2012,kermish_2012},
QUIJOTE~\cite{perez-de-taoro_2014},
SPIDER~\cite{rahlin_2014}, and
SPTPol~\cite{benson_2014,austermann_2012}.
Additional experiments that could use CalSat are being designed:
GLP~\cite{araujo_2014},
SKIP~\cite{johnson_2014}, and
QUBIC~\cite{ghribi_2014}.
And the planned CMB-S4~\cite{abazajian_2015b} program would benefit
from CalSat because it could be used to both accurately calibrate deep
observations and combine results from different observatories.
Though CalSat is designed for CMB polarization experiments, it could
also straightforwardly be used to calibrate other observatories such
as ALMA and the VLA.

%%%%%%%%%%%%%%%%%%%%%%%%%%%%%%%%%%%%%%%%%%%%%%%%%%%%%%%%%%%%%%%%%%%%%%%%%%%%

\section{Discussion}
\label{sec:discussion}

%%%%%%%%%%%%%%%%%%%%%%%%%%%%%%%%%%%%%%%%%%%%%%%%%%%%%%%%%%%%%%%%%%%%%%%%%%%%

As a narrow-band source, CalSat is not designed to characterize the
broad-band spectral properties of CMB polarimeters.
Instead it is designed to provide the critical relationship between
the coordinate system of the polarimeter in the instrument frame and
the coordinate system on the sky that defines the astrophysical $Q$
and $U$ Stokes parameters.
The spectral bands in CMB polarimeters are typically broad, and some
polarimeter technologies, such as the achromatic half-wave plate and
the sinuous antenna multi-chroic pixel, have frequency dependent
performance.
Therefore polarimeter calibration must be performed over the full
spectral bandwidth for experiments using these technologies.
CMB experiment teams must characterize the spectral properties of
their instruments in the laboratory before deployment and then use the
clean and simple CalSat signals together with their lab-based
instrument transfer functions during their calibration analyses.
For example Figure~\ref{fig:hwp_axis} shows simulated modulation
efficiency and polarimeter axis angle curves for an achromatic
half-wave plate polarimeter.
The 80.0, 140 and 249~GHz CalSat tones could be used to verify that the
modulation efficiency is very close to unity at these frequencies and
the polarimeter axis varies as expected as a function of frequency.

CalSat is programmed to listen for the ground station each orbit.
If the radio uplink from the ground station fails, then the on-board
computer turns the CalSat tones off and wait for the link to be
restored.
This mode of operation guards against out-of-control behavior.
If this program malfunctions and the tones stay on, then the batteries
will drain and the instrument will shutdown because the power budget
can not support continuous operation.
Therefore, the most likely failure mode is CalSat would turn off
within 24 hours and then de-orbit 16 years later.

The selected tones match the Amateur Satellite Service Bands as stated
in the text.
We chose to use these bands because there is already an international
agreement in place that allows instruments like CalSat to broadcast in
these bands.
It is possible to request permission to broadcast using other
frequencies for a limited time by applying for an ``experimental''
license from the FCC\footnote{FCC Public Notice DA:13-445}.
We avoided using this approach as a baseline plan because there is
some risk involved.
Projects using experimental licenses cannot cause interference and
they cannot request protection from interference.
The aforementioned Table of Frequency Allocations shows that below
300~GHz, all frequencies are already allocated, so interference is
possible.
Nevertheless, if CalSat moves forward, every effort will be made to
maximize the utility of the instrument by adjusting the frequencies so
they accommodate all possible users.

All experiments trying to extract cosmological information from the
TB, EB and B-mode signals need a robust polarimeter calibration
program that will allow the effect of instrument-induced errors to be
mitigated during data analysis.
A robust mitigation strategy for IPR, which is one of the most
critical systematic errors, has not yet been identified because a
suitable celestial calibration source does not exist and ground-based
solutions are challenging.
Moreover, the commonly used self-calibration technique, that uses the
TB and EB spectra to remove any spurious B-mode signals, prohibits
B-mode measurements from constraining the aforementioned isotropic
departures from the standard model.
CalSat was designed to be a low-cost, open-access solution to the IPR
calibration problem for the CMB community, and its global visibility
makes CalSat the only source that can be observed by all terrestrial
and sub-orbital experiments.
This global visibility makes CalSat a powerful universal standard that
permits comparison between experiments from different observatories
using appreciably different measurement approaches.

%%%%%%%%%%%%%%%%%%%%%%%%%%%%%%%%%%%%%%%%%%%%%%%%%%%%%%%%%%%%%%%%%%%%%%%%%%%%

\section{Acknowledgements}
\label{sec:acknowkedgements}

%%%%%%%%%%%%%%%%%%%%%%%%%%%%%%%%%%%%%%%%%%%%%%%%%%%%%%%%%%%%%%%%%%%%%%%%%%%%

Johnson, Keating and Kaufman acknowledge support from winning a
Buchalter Cosmology Prize (Second Prize) in 2014 for their paper
entitled ``Precision Tests of Parity Violation Over Cosmological
Distances''~\cite{kaufman_2014}.
This paper explores the idea of using TB and EB spectra to study new
physics via CPR, and these kinds of measurements rely on precise
calibration enhancements; CalSat was used as an example of an enhanced
calibration technique in this paper.
We would like to thank Ari Buchalter and the Buchalter Cosmology Prize
Advisory Board and Judging Panel for acknowledging our work with this
prize.

%%%%%%%%%%%%%%%%%%%%%%%%%%%%%%%%%%%%%%%%%%%%%%%%%%%%%%%%%%%%%%%%%%%%%%%%%%%%

%%%%%%%%%%%%%%%%%%%%%%%%%%%%%%%%%%%%%%%%%%%%%%%%%%%%%%%%%%%%%%%%%%%%%%%%%%%%

\end{document}